\documentclass[11pt,aps,nofootinbib,superscriptaddress,amssymb]{revtex4}

\usepackage{amsmath,setspace,subfigure,amsfonts,latexsym}
\usepackage{amssymb}
\usepackage{color}
\usepackage{epsfig}
\usepackage{color}
\usepackage{hyperref}
\usepackage[compat=1.1.0]{tikz-feynman}
\tikzfeynmanset{
every edge={very thick},
}

\definecolor{White}{rgb}{1,1,1}
\definecolor{Red}{rgb}{1,0.1,0}
\definecolor{LightYellow}{rgb}{1,1,.875}
\definecolor{SteelBlue}{rgb}{.273,.508,.703}
\definecolor{navy}{rgb}{0,0,.5}
\definecolor{LightCyan}{rgb}{.875,1,1}
\definecolor{DarkRed}{rgb}{.543,0,0}
\definecolor{HotPink}{rgb}{1,.41,.70}
\definecolor{ForestGreen}{rgb}{.13,.54,.13}
\definecolor{OliveDrab}{rgb}{.42,.55,.14}
\definecolor{MediumBlue}{rgb}{0,0,.80}
\definecolor{RoyalBlue}{rgb}{.25,.41,.88}
\definecolor{DeepSkyBlue}{rgb}{0,.746,1}
\definecolor{Brown}{rgb}{0.545,0.271,0.074}

\def\bec{\begin{center}}
\def\ec{\end{center}}

\def\beq{\begin{equation}}
\def\eeq{\end{equation}}

\newcommand\lsim{\mathrel{\rlap{\lower4pt\hbox{\hskip1pt$\sim$}}
    \raise1pt\hbox{$<$}}}
\newcommand\gsim{\mathrel{\rlap{\lower4pt\hbox{\hskip1pt$\sim$}}
    \raise1pt\hbox{$>$}}}
\def\bea{\begin{eqnarray}}
\def\eea{\end{eqnarray}}
\def\ba{\begin{array}}
\def\ea{\end{array}}

\newcommand{\axino}{{\tilde a}}
\newcommand\unit[1]{\,{\rm #1}}
\newcommand\eV{\unit{eV}}
\newcommand\keV{\unit{keV}}
\newcommand\MeV{\unit{MeV}}
\newcommand\GeV{\unit{GeV}}
\newcommand\TeV{\unit{TeV}}

\begin{document}

\hfill CTPU-17-25 

\hfill UT-17-25

\title{\Large Light Axinos from Freeze-in: 
production processes, phase space distributions, and Ly-$\alpha$ forest constraints}
\author{Kyu Jung Bae} 
\author{Ayuki Kamada} 
\affiliation{Center for Theoretical Physics of the Universe, Institute for Basic Science (IBS), Daejeon 34051, Korea}
\author{Seng Pei Liew}
\affiliation{Physik-Department, Technische Universit\"at M\"unchen, 85748 Garching, Germany}
\author{Keisuke Yanagi}
\affiliation{Department of Physics, University of Tokyo, Bunkyo-ku, Tokyo 113-0033, Japan}

\date{\today}

\begin{abstract}
We consider freeze-in production of $7 \keV$ axino dark matter (DM) in the supersymmetric Dine-Fischler-Srednicki-Zhitnitsky (DFSZ) model in light of the  $3.5 \keV$ line excess.
The {\it warmness} of such $7 \keV$ DM produced from the thermal bath, in general, appears in tension with Ly-$\alpha$ forest data, although a direct comparison is not straightforward.
This is because the Ly-$\alpha$ forest constraints are usually reported on the mass of the conventional warm dark matter (WDM), where large entropy production is implicitly assumed to occur in the thermal bath after WDM particles decouple.
The phase space distribution of freeze-in axino DM varies depending on production processes and axino DM may alleviate the tension with the tight Ly-$\alpha$ forest constraints.
By solving the Boltzmann equation, we first obtain the resultant phase space distribution of axinos produced by 2-body decay, 3-body decay, and 2-to-2 scattering, respectively. 
The reduced collision term and resultant phase space distribution are useful for studying other freeze-in scenarios as well.
We then calculate the resultant linear matter power spectra for such axino DM and directly compare them with the linear matter power spectra for the conventional WDM.
In order to demonstrate realistic axino DM production, we consider benchmark points with the Higgsino next-to-lightest supersymmetric particle (NLSP) and wino NLSP.
In the case of the Higgsino NLSP, the phase space distribution of axinos is colder than that in the conventional WDM case, so the most stringent Ly-$\alpha$ forest constraint can be evaded with mild entropy production from saxion decay inherent in the supersymmetric DFSZ axion model.
\end{abstract}

\maketitle

\titlepage

\tableofcontents

\newpage

\section{Introduction}
\label{sec:intro}

The matter content of the Universe is dominated by unknown particles, which are called dark matter (DM; see Ref.~\cite{Bertone:2016nfn} for a recent historical review).
From its gravitational interaction, we know that such DM consists of non-luminous massive particles.
The microscopic nature of a DM particle may be hinted by its rare decay into a standard model (SM) particle (see Ref.~\cite{Ibarra:2013cra} for a general review).
Actually an unidentified $3.5 \keV$ line has been reported in the X-ray spectra from independent astrophysical objects such as the Perseus galaxy cluster and the Andromeda galaxy, and in independent facilities such as \texttt{Chandra} and \texttt{XMM-Newton} (see Refs.~\cite{Bulbul:2014sua, Boyarsky:2014jta} for the first two reports and also Refs.~\cite{Boyarsky:2014ska, Anderson:2014tza, Iakubovskyi:2015dna, Cappelluti:2017ywp} for following reports).
Perhaps the most popular interpretation of the signal is that it originates from radiative decay of a $7 \keV$ DM particle.%
\footnote{The DM origin of the $3.5 \keV$ line has been challenged by the consistency checks (see Refs.~\cite{Horiuchi:2013noa, Jeltema:2014qfa, Malyshev:2014xqa, Tamura:2014mta, Sekiya:2015jsa, Aharonian:2016gzq} for line searches in different objects or in different instruments and Refs.~\cite{Urban:2014yda, Carlson:2014lla} for a morphological test). 
However, it appears that these constraints are not conclusive (see Refs.~\cite{Boyarsky:2014paa, Bulbul:2014ala, Franse:2016dln} for debates about some of the above constraints), and there is still room for the decaying DM explanation of the $3.5 \keV$ line (see Refs.~\cite{Iakubovskyi:2015wma, Abazajian:2017tcc} for a summary of the current status).
X-ray microcalorimeter sounding rockets may provide a significant test on the DM origin of the $3.5 \keV$ line in near future~\cite{Figueroa-Feliciano:2015gwa}.}
A lot of particle physics models have been suggested to provide a radiatively decaying $7 \keV$ DM particle.

On the other hand, one important point seems having been overlooked: the {\it warmness} of DM.
When keV-scale (or lighter) DM particles are produced from the thermal bath (not necessarily thermally equilibrated), in general, 
a sizable velocity of DM particles affects the evolution of primordial density perturbations and leaves observable signatures on the resultant matter distribution of the Universe.
Such warmness of DM are constrained, for example, by Ly-$\alpha$ forest data.
The latest and strongest constraint is $m_{\rm WDM} \gtrsim 5.3 \keV$~\cite{Irsic:2017ixq} in the conventional warm dark matter (WDM)%
\footnote{The conventional WDM is also referred to as {\it early decoupled thermal relics}.} 
such as light gravitinos in gauge-mediated SUSY breaking scenarios~\cite{Moroi:1993mb, Pierpaoli:1997im}, which are produced and thermalized just after the reheating.
Although the Ly-$\alpha$ forest constraint apparently seems to allow $7 \keV$ DM, we need to remark that they require a very low DM temperature as we will take a closer look in Sec.~\ref{sec:intro_wdm}.
For $7 \keV$ WDM, DM particles need to decouple when the number of effective massless degrees of freedom is as large as $g_{*, {\rm WDM}} \sim 7000$~\cite{Bae:2017tqn} (see Eq.~\eqref{eq:wdm}), which requires further entropy production in addition to the full SM degrees of freedom ($g_{*} = 106.75$).
Since the DM {\it temperature} in other $7 \keV$ DM models (see Eq.~\eqref{eq:nonthermal-temp}) is higher than that in the conventional WDM, the resultant lower bound on the WDM mass is larger; $m = 7 \keV \left( m_{\rm WDM} / 2.5 \keV / ({\tilde \sigma} / 3.6)^{-3/4} \right)^{4/3}$ (see Eq.~\eqref{eq:mass-rel}) when DM particles decouple before the electroweak phase transition.
The Ly-$\alpha$ forest constraint of $m_{\rm WDM} \gtrsim 5.3 \keV$ disfavors such $7 \keV$ DM.

In addition to the temperature, the phase space distribution also affects the warmness of DM through its dimensionless divergence ${\tilde \sigma}^{2}$ (see Eq.~\eqref{eq:sigma}) in the above formula, where ${\tilde \sigma} = 3.6$ is the reference value for the Fermi-Dirac distribution of the conventional WDM.
The non-thermal phase space distributions (especially of sterile neutrino DM) are calculated in the literature~\cite{Boyanovsky:2008nc, Merle:2015oja, Adulpravitchai:2015mna, Venumadhav:2015pla, McDonald:2015ljz, Roland:2016gli, Heeck:2017xbu}.
It is shown that the resultant phase space distribution of {\it freeze-in} DM~\cite{Hall:2009bx} (see also a recent review~\cite{Bernal:2017kxu}) is typically {\it colder} -- higher population at low momenta and thus smaller value of ${\tilde \sigma}$-- than the Fermi-Dirac distribution of the conventional WDM.
The resultant linear matter power spectra, on the other hand, are presented only in limited cases~\cite{Merle:2014xpa, Schneider:2016uqi, Konig:2016dzg}, although a direct comparison of the spectra between non-thermal DM and the conventional WDM models provides a more robust way to convert the Ly-$\alpha$ forest and also other lower bound on the conventional WDM mass into that on the mass of non-thermal DM~\cite{Murgia:2017lwo}.
We demonstrate such a direct comparison by taking $7 \keV$ freeze-in axino DM~\cite{Bae:2017tqn} as an example.
In this paper, furthermore, we explain how we can obtain the phase space distribution from the collision term in a self-contained manner.
For example, we present how we can reduce the collision term into a simple form.
We stress that such methodologies are easily applicable to other models.

Among various attractive particles, the axino, which is the fermion SUSY partner of the axion, is one of the best DM candidates.
For solutions to the gauge hierarchy and strong $CP$ problems, it is plausible to introduce, respectively, supersymmetry (SUSY)~\cite{Nilles:1983ge, Haber:1984rc, Martin:1997ns} and Peccei-Quinn (PQ) symmetry~\cite{Peccei:1977hh, Peccei:1977ur, Weinberg:1977ma, Wilczek:1977pj}.
Consequently, the light axion from PQ symmetry breaking and its SUSY partners are introduced in the model.
In the SUSY limit, the axino is massless since it is a SUSY partner of the massless axion.
Once the SUSY is broken, however, the axino obtains its mass via communication with the SUSY breaking sector.
Although the axino mass is typically of order the gravitino mass, it can be much smaller than the gravitino mass in some models~\cite{Goto:1991gq, Chun:1992zk, Chun:1995hc}.%
\footnote{It is also possible that the axino is much heavier than the gravitino~\cite{Chun:1995hc, Bae:2014efa}.}
In this regard, the axino can be as light as keV and thus it can be a good $7 \keV$ DM candidate.

We introduce R-parity violation to explain the $3.5 \keV$ line excess by axino DM decay.
Such R-parity violation (RPV) may induce harmful proton decay.
However, if the R-parity is violated only in the lepton number violating operators, it retains the proton stability because proton decay requires both lepton and baryon number violation.
In the literature, bilinear and trilinear operators with lepton number violation have been considered for axino decay.
In the case of bilinear R-parity violation (bRPV) in the SUSY Kim-Shifman-Vainshtein-Zakharov (KSVZ) model~\cite{Kim:1979if, Shifman:1979if}, the $3.5 \keV$ line excess requires either a small PQ scale ($v_{\rm PQ} \sim 10^{8} \GeV$)~\cite{Kong:2014gea} or the light Bino ($\sim10 \GeV$)~\cite{Choi:2014tva}.
In the case of trilinear RPV in the SUSY Dine-Fischler-Srednicki-Zhitnitsky (DFSZ) model~\cite{Zhitnitsky:1980tq, Dine:1981rt}, the light stau is necessary to mediate a sufficient axino decay width for the $3.5 \keV$ signal~\cite{Liew:2014gia}.
On the other hand, bilinear RPV in the SUSY DFSZ model induces {\it direct mixings between the axino and neutrinos}, so the axino can decay into light active neutrinos via the mixing~\cite{Chun:1999kd, Choi:2001cm, Chun:2006ss}.
Therefore, as in the case of sterile neutrino decay~\cite{Bulbul:2014sua, Boyarsky:2014jta, Abazajian:2017tcc}, axino decay is able to explain the $3.5 \keV$ signal if the axino-neutrino mixing is realized with $\sin^{2} 2\theta\sim 10^{-10}$ where $\theta$ is the mixing angle~\cite{Bae:2017tqn}.

Production of axinos depends on how the axion supermultiplet interacts with the visible sector, {\it i.e.}, the minimal supersymmetric standard model (MSSM) particles.
In the KSVZ model, the axino couples to the gauginos and gauge bosons via dimension-5 operators, so its production is enhanced at a high temperature~\cite{Covi:2001nw, Brandenburg:2004du, Strumia:2010aa}.
On the other hand, in the DFSZ model, the axino couples to the Higgses and Higgsino via effectively dimension-4 operators.
The dimension-5 operator of axino-gaugino-gauge boson couplings is also generated as in the KSVZ model. 
In the DFSZ model, however, the dimension-5 operator is suppressed at the scale above the $\mu$-term, so the effect is negligible even if the reheating temperature is very high~\cite{Bae:2011jb}.
Due to its apparently renormalizable couplings in the DFSZ model, axinos are dominantly produced near the threshold scale of the process ({\it e.g.}, $\mu$ for Higgsino decay into the axino)~\cite{Bae:2011jb,Chun:2011zd,Bae:2011iw}
and thus it shows the freeze-in nature of feebly interacting particles.

The saxion, which is the scalar SUSY partner of the axion, is also an important ingredient for the abundance and phase space distribution of axinos.
While the saxion abundance from thermal production is similar to that of the axino, the saxion abundance from the coherent oscillation can be much larger and thus dominate the Universe.
If the coherent oscillation of the saxion dominates the Universe and decays after axino production, it releases a certain amount of entropy.
Consequently, axinos produced before saxion decay are diluted, and also their momenta are redshifted ({\it i.e.}, become colder) by the entropy.
Entropy production from saxion decay may be crucial for $7 \keV$ axino DM since the strongest constraint from Ly-$\alpha$ forest data has a tension even with freeze-in production when we consider realistic models~\cite{Bae:2017tqn}.
We also discuss how we can infer the required entropy dilution factor to evade the constraint.
This can be done based on a simple extension of the characteristic velocity, which will be introduced around Eq.~\eqref{eq:sec_moment}.

The paper is organized as follows.
The SUSY DFSZ model is described in Sec.~\ref{sec:axino-model}, where we introduce the R-parity violation and show that decaying $7 \keV$ axino DM behaves similarly to sterile neutrino DM in regard to the $3.5 \keV$ line excess.
The freeze-in production channels of the axino and dilution of its abundance from saxion domination and subsequent decay are also explained.
We take a closer look at the tension between $3.5 \keV$ line-motivated $7 \keV$ DM and the Ly-$\alpha$ forest constraints in Sec.~\ref{sec:intro_wdm}.
In Sec.~\ref{sec:mom-dist-axino}, we first focus on each production channel and reduce the collision term in the Boltzmann equation to a simple form, while devoting appendix~\ref{sec:boltzmann-equation} to the details.
The reduced Boltzmann equation is numerically integrated with the squared matrix elements given in appendix~\ref{sec:matrix-elements}, and the resultant axino phase space distributions are compared among different production channels.
A simple fitting function with two parameters for phase space distributions are also provided.
Next, we introduce more realistic scenarios, where several production channels simultaneously contribute to the axino abundance, and thus the resultant axino distribution is given by a yield-weighted {\it superposition} of those from each channel.
The fitting parameters of the realistic axino distributions are summarized in appendix~\ref{sec:fitt-fuct-benchm}.
In Sec.~\ref{sec:mps}, by using the obtained axino distributions, we follow evolution of the primordial density perturbations.
The resultant matter power spectra are compared to those in the conventional WDM.
From the comparison, we infer the required entropy dilution factor from saxion decay to evade the most stringent Ly-$\alpha$ forest constraint.
Furthermore, by directly comparing the resultant matter power spectra with those in the conventional WDM, we confirm that $7 \keV$ axino DM with the inferred entropy dilution factor is viable in regard to the Ly-$\alpha$ forest constraint.
We also discuss the possibility that the Ly-$\alpha$ forest constraints are evaded by a compressed mass spectrum.
Sec.~\ref{sec:concl} is devoted to concluding remarks.

\section{Model of the axino with R-parity violation}
\label{sec:axino-model}

We discuss the SUSY DFSZ axion model to describe light axino decay and production.
The relevant interactions in this model are obtained from higher dimensional operators and PQ symmetry breaking.
The $\mu$-term is generated by the Kim-Nilles mechanism~\cite{Kim:1983dt}.
On production of axinos, the $\mu$-term interaction generates main production processes including 2-to-2 scattering and heavy particle decay.
Such production processes determine not only the total abundance of axinos but also their phase space distribution, since axinos have highly feeble interactions so that the phase space distribution is not re-distributed but just redshifted by the cosmic expansion after they are produced.
In addition, bRPV terms are naturally introduced in the same manner, while it is even more suppressed than the $\mu$-term by a proper PQ charge assignment.
One finds an axino-neutrino mixing in this model and thus explains axino decay in the same way as sterile neutrino decay.

\subsection{DFSZ model with bRPV}
\label{sec:DFSZwbRPV}

In the DFSZ axion model, the Higgs doublets ($H_{u}$ and $H_{d}$) are charged under PQ symmetry, so the bare mass term of $H_{u} H_{d}$ is generated by PQ symmetry breaking.
In the MSSM, such a term is the $\mu$-term, and is given by the superpotential,
\begin{equation}
  W_{\rm DFSZ} = \lambda Z \left(X Y - \frac{v_{\rm PQ}^{2}}{2} \right) + \frac{y_{0}}{M_{*}} X^{2} H_{u} H_{d} \,,
\end{equation}
where $X$, $Y$, and $Z$ are chiral superfields with respective PQ charges being $Q_{\rm PQ}\{Z, X, Y, H_{u}, H_{d}\} = \{0, -1, 1, 1, 1\}$.
The dimensionless coupling constants are denoted by $\lambda$ and $y_0$, while $M_{*}$ is the scale of UV physics.
Once PQ symmetry is broken, $\langle X \rangle \sim \langle Y \rangle \sim v_{\rm PQ} / \sqrt{2}$, the $\mu$-term is generated,
\begin{equation}
  \mu \sim \frac{y_{0}v_{\rm PQ}^{2}}{2M_{*}} \,.
\end{equation}
When $M_{*} \sim 10^{16} \GeV$, $y_{0} \sim 0.1$, and $v_{\rm PQ} \sim 10^{10} \GeV$, one obtains $\mu \sim 500 \GeV$.
This is a solution to the SUSY $\mu$-term problem via the Kim-Nilles mechanism~\cite{Kim:1983dt}.
In addition, bRPV terms can also be introduced,
\begin{equation}
  W_{\rm bRPV} = \frac{y_{i}'}{M_{*}^{2}} X^{3} L_{i} H_{u} \,,
\end{equation}
where $L_{i}$ [$i (=1 \text{--} 3$) denotes the flavor] is the lepton doublet with $Q_{\rm PQ}(L_{i}) = 2$.
From PQ symmetry breaking, one finds
\begin{equation}
  \mu_{i}' \sim \frac{y_{i}' v_{\rm PQ}^{3}}{2 \sqrt{2} M_{*}^{2}} \,.
\end{equation}
As for the $\mu$-term generation, when $M_{*} \sim 10^{16} \GeV$, $y_{i}'\sim 1$ and $v_{\rm PQ} \sim 10^{10} \GeV$, one obtains a tiny bRPV interaction as $\mu_{i}' \sim 10^{-3} \GeV$.

When PQ symmetry is broken, PQ fields can be expressed as
\begin{equation}
  X = \frac{v_{\rm PQ}}{\sqrt{2}} e^{A / v_{\rm PQ}} \,, \qquad 
  Y = \frac{v_{\rm PQ}}{\sqrt{2}} e^{-A / v_{\rm PQ}} \,.
\end{equation}
The axion superfield ($A$) consists of the axion ($a$), saxion ($s$), and axino ($\axino$),
\begin{equation}
  A = \frac{s + i a}{\sqrt{2}} + \sqrt{2} \theta\axino + \theta^{2}{\cal F}_{A} \,,
\end{equation}
where $\theta$ is the superspace coordinate and ${\cal F}_{A}$ is the $F$-term of the axion superfield.
One obtains an effective superpotential,
\begin{eqnarray}
  W_{\rm eff}
  &=& \mu e^{c_{H} A / v_{\rm PQ}} H_{u} H_{d} + \epsilon_{i}\mu e^{c_{p} A / v_{\rm PQ}} L_{i} H_{u} \notag \\
  &\simeq& \mu \left( 1 + \frac{c_{H} A}{v_{\rm PQ}} \right) H_{u} H_{d} + \epsilon_{i}\mu \left( 1 + \frac{c_{p} A}{v_{\rm PQ}} \right) L_{i} H_{u} \,,
  \label{eq:DFSZlag}
\end{eqnarray}
where $\epsilon_{i} = \mu_{i}' / \mu$, $c_{H} = 2$, and $c_{p} = 3$.
The approximation in the second line is valid for evaluating fermion masses and mixings.
From this superpotential, one can easily find the mixing angle between the axino and active neutrinos, which is given by
\begin{equation}
  |\theta| 
  \simeq \frac{\epsilon \mu v_{u} (c_{p} - c_{H})}{m_{\axino} v_{\rm PQ}}
  \simeq 10^{-5} \left( \frac{\epsilon}{10^{-5}} \right) \left( \frac{\mu}{400 \GeV} \right) \left( \frac{7 \keV}{m_{\axino}} \right) \left(\frac{10^{10} \GeV}{v_{\rm PQ}}\right) \,,
\end{equation}
where we assume that a sizable bRPV exists in only the third generation, {\it i.e.}, only $\epsilon_{3}$ is relevant and others are negligible. We also simply write $\epsilon_{3} = \epsilon$.
One can easily find the proper axino-neutrino mixing, $\sin^{2} 2 \theta \sim 10^{-10}$, which explains the $3.5 \keV$ line excess as in the case of sterile neutrino DM.%
\footnote{It is also noted that the sneutrino obtains a non-zero vacuum expectation value (VEV) $\langle {\tilde \nu} \rangle$ due to the bRPV scalar potential $- {\cal L}_{\rm RPV} = B_{i} {\tilde L}_{i} H_{u} + m_{L_{i} H_{d}}^{2} \tilde L_{i} H_{d}^{\dagger} + {\rm h.c.}$ 
This induces mixings between the neutrinos and gauginos, mediating axino decay via the axino-photino-photon effective operator~\cite{Kim:2001sh, Chun:2006ss, Endo:2013si, Kong:2014gea, Choi:2014tva}, as well as lepton-flavor violation processes and neutrino masses. Among them, neutrino masses impose the most stringent bound ($\langle {\tilde \nu} \rangle/v \lesssim 10^{-7}$ for the gaugino masses of $O(100)$\,GeV)~\cite{Barbier:2004ez}.
This bound is avoided by assuming that the sneutrino VEV is sufficiently small (this is achievable by,  {\it e.g.}, raising the sneutrino mass).}

With this mixing, however, the axino (as the sterile neutrino) abundance from the Dodelson-Widrow mechanism~\cite{Dodelson:1993je} is an order of magnitude smaller than the total DM abundance~\cite{Abazajian:2005gj, Asaka:2006nq},
\begin{equation}
\Omega_{\axino} h^{2} \simeq 0.01 \left( \frac{\sin^{2} 2 \theta}{2 \times 10^{-10}} \right) \left( \frac{m_{\axino}}{7 \keV} \right)^{2} \,.
\end{equation}
Although the preferred value of the mixing angle for the $3.5 \keV$ line excess varies depending mainly on the objects~\cite{Bulbul:2014sua, Boyarsky:2014jta, Abazajian:2017tcc}, $\sin^{2} 2 \theta \simeq 1 \text{--} 4 \times 10^{-11}$ seems viable once the constraints from the \texttt{Chandra} observation of the Andromeda galaxy~\cite{Horiuchi:2013noa} and the \texttt{XMM-Newton} observation of dwarf spheroidal galaxies~\cite{Malyshev:2014xqa} are taken into account.
With such a mixing angle, axinos produced from the Dodelson-Widrow mechanism account for at most a few $\%$ of the whole DM abundance.
Therefore, additional production processes must exist for the DM abundance.
In the DFSZ axino case, interactions from the $\mu$-term dominantly produce axinos via the freeze-in mechanism.
We will discuss axino production in more detail in the following subsection.

\subsection{Axino production}
\label{sec:axinoprod}

In the case of DFSZ axino, dominant production processes are generated by the interaction accompanying the $\mu$-term in Eq.~\eqref{eq:DFSZlag}.
The bRPV term is suppressed by ${\cal O}(v_{\rm PQ}/M_{*})$ compared to the $\mu$-term so that the contribution is always negligible.
Since PQ symmetry is subject to the quantum anomaly, there is an axion-gauge-gauge interaction that is necessary for a solution to the strong $CP$ problem.
By field rotation of the Higgses and quarks, one can find the interaction,
\begin{equation}
  {\cal L} \supset - \frac{g_{s}^{2}}{32 \pi^{2} v_{\rm PQ} / N_{\rm DW}} \int d^{2}\theta A W^{b} W^{b} + {\rm h.c.} \,,
  \label{eq:aGG}
\end{equation}
where $W^{b}$ is the gauge superfield for gluons, $g_{s}$ is the strong coupling constant, and $N_{\rm DW} = 6$ is the domain wall number.
It apparently seems that Eq.~\eqref{eq:aGG} can contribute to axino production at a high temperature since it is a non-renormalizable interaction.
However, as argued in Ref.~\cite{Bae:2011jb}, the one-particle-irreducible amplitude for axino-gluino-gluon is suppressed at the energy above the scale of $\mu$ so that we can safely neglect this operator in DFSZ axino production.
Therefore, from now on, we will consider only the $\mu$-term interaction in Eq.~\eqref{eq:DFSZlag} for axino production.

Since the interaction in Eq.~(\ref{eq:DFSZlag}) generates a dimensionless coupling constant, the scattering cross section corresponding to axino production increases as the temperature decreases.
For this reason, the scattering rate per Hubble time is maximized at the lowest possible temperature.
If the temperature is lower than the threshold scale of the process, the reaction rate is suppressed by the Boltzmann factor and thus the process becomes irrelevant.
Therefore dominant axino production occurs at the temperature near the threshold scale. 
Note that axinos are never thermalized while the other particles in the process are in thermal equilibrium.

In the SUSY limit, the axino yield from scattering with gauge particles is given by~\cite{Bae:2011jb}
\begin{equation}
  Y_{\axino}^{\rm scat} \simeq \frac{3 c_{H}^{2}}{2} \frac{135 \sqrt{10}}{4 \pi^7 g_{*}^{3/2}} \frac{g_{2}^{2} M_{P} \mu}{v_{\rm PQ}^{2}} \,,
\end{equation}
where $g_{2}$ is the SU(2) gauge coupling constant and $M_{P}$ is the reduced Planck scale.
A similar contribution can also be obtained from top Yukawa interactions~\cite{Bae:2011iw}, which is larger and thus taken into account in the following section.
This yield does not depend on the reheating temperature $T_{R}$ as long as $T_{R} > \mu$ is retained.
More interestingly, it is worth noting that $Y_{\axino}^{\rm scat}$ is proportional to $\mu$.
This is because the axino-Higgs-Higgsino coupling constant, $\mu / v_{\rm PQ}$, is enhanced for large $\mu$ although the threshold scale (Higgsino mass) suppresses the overall yield by a factor of $(1 / \mu)$.
In the broken SUSY case, on the other hand, the soft terms for the gaugino and Higgs masses must be taken into account.
Due to the heavy gauginos and scalars, some possible channels are highly suppressed so that the axino yield from these processes can alter.
Furthermore, the phase space distribution can differ for each production channel.

As we consider axino DM, heavy particle decay is also important for axino production.
In particular, 2-body decay of the Higgses or Higgsino significantly contributes if the Higgsino is light compared to other SUSY particles.
The axino yield from decay is given by~\cite{Chun:2011zd},
\begin{equation}
  Y_{\axino}^{\rm decay} \simeq \frac{135 c}{4 \pi^{4} g_{*}^{3/2}} \frac{\Gamma M_{P}}{m^{2}} \int z^{3} K_{1}(z) dz \,,
\end{equation}
where $K_{1}$ is the 1st-order modified Bessel function and $z = m / T$.
The $\Gamma$ and $m$ are, respectively, the decay width and mass of the decaying particle (Higgs or Higgsino).
The number of degrees of freedom of the process is denoted by $c$. 
If the Higgsino decays into the axino and lighter Higgs, the decay width of ${\widetilde H} \to \axino + H_{L}$ is given by%
\footnote{We define $H_{L}$ ($H_{H}$) as the lighter (heavier) Higgs in the mass eigenstate after diagonalizing the mass matrix of $H_{u}$ and $H_{d}$ (see appendix~\ref{sec:higgses}). In this notation, $H_{L}$ is the SM-like Higgs at the decoupling limit.} 
\begin{equation}
  \Gamma({\widetilde H} \to \axino + H_{L}) = \frac{c_{H}^{2}}{32 \pi} \left( \frac{\mu}{v_{\rm PQ}} \right)^{2} \mu \,.
\end{equation}
The mass and number of degrees of freedom are, respectively, $m = \mu$ and $c = 2$.
On the other hand, if the lighter Higgs decays into the axino and Higgsino, one obtains
\begin{equation}
  \Gamma(H_{L} \to \axino + {\widetilde H}) = \frac{c_{H}^{2}}{16 \pi} \left( \frac{\mu}{v_{\rm PQ}} \right)^{2} m_{H_{L}} \,,
\end{equation}
where $m = m_{H_{L}}$ and $c = 4$.
In the above two formulas, we ignore the masses of the final-state particles.

If the Higgsino is very heavy, more specifically, heavier than the gaugino, 3-body decay of the gaugino can also make a significant contribution to axino production.
For instance, if $M_{2} \ll \mu$ ($M_{2}$ is the wino mass) and $T_{R} < \mu$, axinos are produced by thermal winos through ${\widetilde W} \to \axino + H_{L} + H_{L}^{*}$.
In this case, it is worth noting that 2-to-2 scattering such as ${\widetilde W} + H_{L} \to \axino + H_{L}$ is also comparable.
This is simply understood since the phase space factor of these two processes are of the same order and corresponding Feynman diagrams are related by crossing symmetry.

\subsection{Dilution from saxion decay}
\label{sec:entropyprod}

In the SUSY axion model, saxion production and decay have to be taken into account since its energy density can dominate the Universe,  and its late decay can inject entropy only into the SM sector, and thus it affects the DM abundance and temperature.
In the SUSY DFSZ model, the saxion abundance from thermal production is of the same order as axino thermal production.
However, the saxion can be produced in the form of coherent oscillation; the yield is given by~\cite{Lessa:2011pqa}
\begin{equation}
  Y_{s}^{\rm CO} \simeq 1.9 \times 10^{-6} \left( \frac{\rm GeV}{m_{s}} \right) \left( \frac{{\rm min}[T_{R}, T_{s}]}{10^{7} \GeV} \right) \left( \frac{s_{0}}{10^{12} \GeV} \right)^{2} \,,
\end{equation}
where $s_{0}$ is the initial saxion amplitude when it starts oscillation.
Here the temperature $T_{s}$ is defined by $3 {\dot R} / R (T_{s}) = m_{s}$, where $R$ is the cosmic scale factor and the dot denotes the derivative with respect to the cosmic time $t$.
The saxion can dominate the energy density of the Universe when the temperature becomes the equality temperature, which is given by
\begin{equation}
  T_{e}^{s} 
  = \frac43 m_{s}Y_{s}^{\rm CO}
  \simeq 2.5 \times 10^{2} \GeV \left( \frac{{\rm min}[T_{R},T_{s}]}{10^{7} \GeV} \right) \left( \frac{s_{0}}{10^{16} \GeV} \right)^{2} \,.
  \label{eq:eq_tem}
\end{equation}
Later, the saxion decays at the temperature $T_{D}^{s}$, and it produces an amount of entropy of~\cite{Kolb:1990vq},
\begin{equation}
  \label{eq:dil-fac}
  \Delta \simeq \frac{T_{e}^{s}}{T_{D}^{s}} \,.
\end{equation}
Consequently, the axino abundance is reduced by this dilution factor.%
\footnote{The axino abundance from the Dodelson-Widrow mechanism may not be diluted by saxion domination and subsequent decay because in this case axinos are produced at a low temperature, $T \sim 100 \MeV$~\cite{Dodelson:1993je}. 
Such axinos, on the other hand, account for a few $\%$ of the whole DM abundance as discussed in Sec.~\ref{sec:DFSZwbRPV}.
Therefore, in the following discussion, we ignore axinos produced from the Dodelson-Widrow mechanism.}
Moreover, saxion decay does {\it reheat} the thermal plasma but does not affect the axino temperature, so axinos become much colder than those before saxion decay.
As we will see in Sec.~\ref{sec:mps}, this plays an important role in making axino DM colder and thus concordant with Ly-$\alpha$ forest constraints.

\section{$7 \keV$ DM and Ly-$\alpha$ forest constraints}
\label{sec:intro_wdm}

The warmness of DM is limited by increasingly stringent Ly-$\alpha$ forest constraints.
For instance, the 30 \texttt{HIRES}, 23 \texttt{LRIS}, and 27 \texttt{UVES} high-redshift quasar spectra place a lower bound ($2\sigma$) on the conventional WDM mass as $m_{\rm WDM} \gtrsim 2.0 \keV$~\cite{Viel:2005qj}; the spectra measured by early \texttt{SDSS} place $m_{\rm WDM} \gtrsim 2.5 \keV$~\cite{Seljak:2006qw} ($2 \keV$ in an independent analysis~\cite{Viel:2006kd}); the 55 \texttt{HIRES} spectra combined with the 132 \texttt{SDSS} spectra place $m_{\rm WDM} \gtrsim 4 \keV$~\cite{Viel:2007mv}; the 25 spectra by \texttt{HIRES} and \texttt{MIKE} place $m_{\rm WDM} \gtrsim 3.3 \keV$~\cite{Viel:2013apy}; the 13,821 \texttt{SDSS-III/BOSS} spectra place $m_{\rm WDM} \gtrsim 4.09 \keV$ ($2.96 \keV$ when the Ly-$\alpha$ forest data are combined with the \texttt{Planck} measurement of cosmic microwave background anisotropies)~\cite{Baur:2015jsy}; and the \texttt{XQ-100} spectra combined with the 25 spectra by \texttt{HIRES} and \texttt{MIKE} place $m_{\rm WDM} \gtrsim 5.3 \keV$~\cite{Irsic:2017ixq} ($4.65 \keV$ in an independent analysis where 13,821 \texttt{SDSS-III/BOSS} spectra are added~\cite{Yeche:2017upn}).
These constraints are so strong that they exclude WDM as a solution to the small-scale tensions or leave only a small parameter region if any~\cite{Schneider:2013wwa}.
One may wonder if we can alleviate the tight constraints by considering a mixed DM model, where DM consists of cold and warm components~\cite{Harada:2014lma, Schneider:2014rda, Kamada:2016vsc}.
The recent analysis of Ly-$\alpha$ forest data, on the other hand, may disfavor even a mixed DM model as a solution to the small-scale crisis~\cite{Baur:2017stq}.

It is often not trivial nor direct to convert the constraint on the conventional WDM mass into that on other DM models.
In the conventional WDM model, the DM particles are assumed to follow the Fermi-Dirac distribution with two spin degrees of freedom, where the mass and temperature are parameters: $m_{\rm WDM}$ and $T_{\rm WDM}$.
The temperature is fixed to reproduce the observed relic density of DM for a given mass.
The WDM density parameter is given by 
\begin{equation}
  \Omega_{{\rm WDM}, 0} h^{2} = \left( \frac{m_{\rm WDM}}{ {\rm 94 \eV}} \right) \left( \frac{T_{{\rm WDM}, 0}}{T_{\nu, 0}} \right)^{3} = 7.5 \left(\frac{m_{\rm WDM}}{7 \keV}\right) \left(\frac{106.75}{g_{*, {\rm WDM}}}\right) \,, 
  \label{eq:wdm}
\end{equation}
where $h$ is the dimensionless Hubble constant and $T_{\nu}$ is the temperature of the SM neutrino, and the subscript $0$ means that the quantity is evaluated at present.
In the last equality, we use $T_{{\rm WDM}, 0} = (10.75 / g_{*, {\rm WDM}})^{1/3} T_{\nu, 0}$, which is derived from the comoving entropy conservation.
This shows that we need $g_{*, {\rm WDM}} \sim 7000$ although $g_{*} = 106.75$ (226.75) even with full SM (MSSM) degrees of freedom.
It implies that a large entropy dilution factor, $\Delta \sim 70$, is needed after WDM decoupling.

Then a question is what the lower bound on the DM mass is when the DM particles are produced and decoupled before the electroweak phase transition.
The simplest way to infer a lower bound on the DM mass in a non-conventional model is equating the {\rm naive} velocities in two models, which are defined by the temperature divided by the mass: $\sigma_{\rm naive} = T_{{\rm DM}, 0} / m$.
In the conventional WDM model, this takes a value of $\sigma_{\rm naive} \simeq 3.8 \times 10^{-8} \left( {\rm keV} / m_{\rm WDM} \right)^{4/3}$.
The temperature in non-conventional DM models can be defined similarly to that in the conventional WDM model;
\begin{align}
  \label{eq:nonthermal-temp}
  T_{\rm DM} = \left( \frac{g_{*}(T)}{g_{*}(T_{\rm dec})} \right)^{1/3} T \,,
\end{align}
where $T$ is the temperature of the thermal plasma and $T_{\rm dec}$ is taken as the temperature at the decoupling of DM.
Just after the decoupling, $T_{\rm DM} = T$.
The DM temperature is related with the temperature of the SM neutrino at present as $T_{{\rm DM}, 0} = (10.75/106.75)^{1/3} T_{\nu, 0}$ for DM particles decoupled before the electroweak phase transition, and thus we obtain $\sigma_{\rm naive} \simeq 1.1 \times 10^{-8} \left(7 \keV / m \right)$.
It ends up with the relation of $m = 7 \keV \left( m_{\rm WDM} / 2.5 \keV \right)^{4/3}$.
Therefore, if one takes only a less stringent constraint like $m_{\rm WDM} \gtrsim 2.0 \keV$, $7 \keV$ DM is viable, but if one relies on severer constraints like $m_{\rm WDM} \gtrsim 3.3 \keV$, $7 \keV$ DM is in tension.

In the above discussion one may wonder how we can define the temperature in general for non-thermally distributed DM particles.
To make the discussion broadly applicable, we should quantify the warmness in terms of the phase space distribution.
The following velocity is suggested to characterize well the cutoff scale of the resultant matter power spectrum~\cite{Kamada:2013sh}:
$\sigma = \sqrt{\langle p^{2} \rangle} / m$,
where
\begin{equation}
  \langle p^{2} \rangle =  \frac{1}{n} g \int \frac{d^{3} p}{(2 \pi)^{3}} \, p^{2} f(p) \,,
  \label{eq:sec_moment}
\end{equation}
with $p$ being the absolute value of the three-momentum and $g$ being the number of spin degrees of freedom of the DM particle.
The phase space distribution $f (p)$ is normalized such that the number density is given by $n = g \int d^{3} p / (2 \pi)^{3} f(p)$.
We can relate $\sigma$ and $\sigma_{\rm naive}$ by introducing the comoving momentum, $q = p / T_{\rm DM}$, and
\begin{equation}
  {\tilde \sigma}^{2} = \frac{\int dq q^{4} f(q)}{\int dq q^{2} f(q)} \,.
  \label{eq:sigma}
\end{equation}
One finds $\sigma = {\tilde \sigma} \sigma_{\rm naive}$, and ${\tilde \sigma}$ is particularly useful when we constrain realistic models whose phase space distributions are given by a superposition of those from respective production channels.
This is because ${\tilde \sigma}_{\rm realistic}^{2}$ is written by the $Y$-weighted sum of ${\tilde \sigma}_{\rm ch}^{2}$ in the respective channels as
\begin{align}
  \label{eq:sigma-decmp}
  {\tilde \sigma}_{\rm realistic}^{2}
  = \sum_{\rm channels} \frac{Y_{\rm ch}}{Y_{\rm total}} {\tilde \sigma}_{\rm ch}^{2} \,,
\end{align}
and may be determined by the channel that dominates axino production of a given model.
Now the lower bounds on the thermally distributed (conventional) WDM mass and non-thermally distributed DM mass are related as
\begin{align}
  \label{eq:mass-rel}
  m = 7 \keV \left(\frac{m_{\rm WDM}}{2.5 \keV ({\tilde \sigma} / 3.6)^{-3/4}} \right)^{4/3} \,.
\end{align}
Remark that in the conventional WDM model, $f(q) = 1 / (e^{q} + 1)$ and thus ${\tilde \sigma} = \sqrt{15 \zeta(5) / \zeta(3)} \simeq 3.6$, where $\zeta(x)$ is the Riemann zeta function.
On the other hand, $f(q)$ of non-conventional models generally depends on the production mechanism.
When DM particles are produced by 2-body decay of a heavy particle ({\it i.e.}, the freeze-in mechanism), it is known that the phase space distribution can be approximated by $f(q) \approx q^{- 0.5} e^{- q}$~\cite{Boyanovsky:2008nc}, which leads to ${\tilde \sigma} = \sqrt{35} / 2 \simeq 3.0$ and $2.5 \keV ({\tilde \sigma} / 3.6)^{-3/4} = 2.9 \keV$.
Note that taking account of the non-thermal phase space distribution ends up with a $20 \%$ weaker constraint than that inferred by the naive estimation.
This is because the phase space distribution from 2-body decay of a heavier particle is colder than the Fermi-Dirac one.
This exercise drives us to examine more closely the effects of non-thermal phase space distributions on the resultant matter distribution in the following sections.

\section{Phase space distribution of axinos}
\label{sec:mom-dist-axino}

In this section, we study the phase space distribution of freeze-in axinos.
As discussed in Sec.~\ref{sec:axino-model}, axinos are produced via various freeze-in processes.
Once the SUSY spectrum is fixed, the axino yield and phase space distribution are the sum of the contributions from respective production processes.
The axino phase space distribution takes different forms for various production processes, and the resultant linear matter power spectrum depends on which process is dominant.
Therefore, in order to compare the resultant matter power spectrum to the Ly-$\alpha$ forest constraints, we should clarify the relation between production mechanisms and shapes of the axino phase space distribution.

\subsection{Boltzmann equation}
\label{sec:boltzmann-eq}

Axino production is described by the following Boltzmann equation in the homogeneous and isotropic Universe:
\begin{align}
  \label{eq:boltzmann}
  \frac{df_{\axino}(t, p)}{dt}
  = \frac{\partial f_{\axino}(t, p)}{\partial t} - \frac{{\dot R}(t)}{R(t)} p \frac{\partial f_{\axino}(t, p)}{\partial p}
  = \frac{1}{E_{\axino}} C(t, p) \,,
\end{align}
where $E_{\axino} = \sqrt{m_{\axino}^{2} + p^{2}}$ is the axino energy.
The collision term $C(t, p)$ contains all the interaction for axino production and annihilation.
However, since the axino abundance is small compared to the particles in the thermal plasma, we can neglect $f_{\axino}$ in the collision term.
With $f_{\axino} \ll 1$, the collision term is the sum of the contributions from respective production processes.
A contribution from $1 + 2 + \cdots \to \axino + 3 + 4 + \cdots$ is written as 
\begin{align}
  \label{eq:col-term}
  \frac{g_{\axino}}{E_{\axino}} C_{1 + 2 + \cdots \to \axino + 3 + 4 + \cdots}(t, p_{\axino})
  =&  \frac{1}{2 E_{\axino}} \int \prod_{i \neq \axino} \frac{d^{3} p_{i}}{(2 \pi)^{3} 2 E_{i}} (2 \pi)^{4} \delta^{4}(\hat{p}_{1} + \hat{p}_{2} + \cdots - \hat{p}_{\axino} - \hat{p}_{3} - \hat{p}_{4} - \cdots) \notag \\
  & \times \sum_{\rm spin} |{\cal M}_{1 + 2 + \cdots \to \axino + 3 + 4 + \cdots}|^{2} f_{1} f_{2} \cdots (1 \mp f_{3})(1 \mp f_{4}) \cdots \,,
\end{align}
where $\hat{p}_{i}$ is the four-momentum of particle $i$, $g_{\axino} = 2$, and the spin sum is taken over both initial- and final-state particles.
Since Eq.~\eqref{eq:col-term} is independent of $f_{\axino}$, 
we can simply integrate Eq.~\eqref{eq:boltzmann} to obtain the phase space distribution at later time $t_{f}$:
\begin{align}
  \label{eq:integ-boltz}
  f_{\axino}(t_{f}, p) = \int_{t_{i}}^{t_{f}} dt  \frac{1}{E_{\axino}} C\left( t, \frac{R(t_{f})}{R(t)} p \right) \,,
\end{align}
where $t_{i}$ is the reheating time.
Note that since the momentum is redshifted, the axino with the momentum $p$ at $t = t_{f}$ must have the momentum of $R(t_{f}) / R(t) \, p$ at earlier time $t < t_{f}$.

Before going to specific examples, we give useful formulas of $C(t, p)$.
They are generic and applicable to other models with the axino being replaced by a freeze-in particle of interest.
Once we calculate the squared matrix element of a given process, $\sum_{\rm spin} |{\cal M}|^{2}$, the axino phase space distribution is obtained by Eq.~\eqref{eq:integ-boltz} and the following formulas.
We assume that particles other than the axino are thermally equilibrated.
We neglect the Bose enhancement and Pauli blocking factors of the particles in the thermal plasma, $1 \pm f_{i}^{\rm eq} \simeq 1$, whose effects are so small that the following discussion is not affected (see appendix~\ref{sec:appr-1pm-fsim}).

\subsubsection{2-body decay}
\label{sec:2-body-decay}

For 2-body decay, $1 \to \axino + 2$, the collision term is put into
\begin{align}
  \label{eq:col-2body}
  \frac{g_{\axino}}{E_{\axino}} C_{1 \to \axino + 2}(t, p_{\axino})
  &= \pm\frac{T}{16 \pi p_{\axino} E_{\axino}} \sum_{\rm spin} |{\cal M}_{1 \to \axino + 2}|^{2} \ln{\left( \frac{1 \pm e^{- (E_{2}^{-} + E_{\axino}) / T}}{1 \pm e^{- (E_{2}^{+} + E_{\axino}) / T}} \right)} \,,
\end{align}
where $+$ ($-$) is taken when particle 1 is a fermion (boson).
We define $E_{2}^{\pm} = \sqrt{m_{2}^{2} + (p_{2}^{\pm})^{2}}$ and $p_{2}^{\pm}$ by the solutions of the energy conservation equations:%
\footnote{Note that $+$/$-$ in $E^{\pm}_2$ and $p_2^{\pm}$ are a symbol related to kinematics not to boson/fermion.}
\begin{align}
  \label{eq:2body-epm}
  \sqrt{m_{1}^{2} + (p_{2}^{\pm} \pm p_{\axino})^{2}} = E_{2}^{\pm} + E_{\axino} \,.
\end{align}
This agrees with the result of Ref.~\cite{Boyanovsky:2008nc}.

\subsubsection{Scattering}
\label{sec:scattering}

For 2-to-2 scattering, $1 + 2 \to \axino + 3$, the collision term is put into (see appendix~\ref{sec:derivation-eq})
\begin{align}
  \label{eq:col-scat}
  \frac{g_{\axino}}{E_{\axino}} C_{1 + 2 \to \axino + 3}(t, p_{\axino})
  &=
    \pm\frac{T}{512 \pi^{3} p_{\axino} E_{\axino}} e^{- E_{\axino} / T} \int ds \frac{1}{\sqrt{s}p_{3 \axino}} \ln{\left( \frac{1 \pm e^{- E_{3}^{-}(s) / T}}{1 \pm e^{- E_{3}^{+}(s) / T}} \right)} \int dt \sum_{\rm spin} |{\cal M}_{1 + 2 \to \axino + 3}|^{2} \,,
\end{align}
where $+$ ($-$) sign is taken when particle 3 is a fermion (boson).
The kinematic variables are defined by
\begin{align}
  s &= (p_{1} + p_{2})^{2} = (p_{\axino} + p_{3})^{2} \,, \label{eq:def-s} \\
  t &= (p_{1} - p_{\axino})^{2} = (p_{2} - p_{3})^{2} \,, \\
  u &= (p_{1} - p_{3})^{2} = (p_{2} - p_{\axino})^{2} = m_{1}^{2} + m_{2}^{2} + m_{3}^{2} + m_{\axino}^{2} - s - t \,, \label{eq:def-u} \\
  p_{3 \axino} &= \frac{\sqrt{s - (m_{\axino} + m_{3})^{2}} \sqrt{s - (m_{\axino} - m_{3})^{2}}}{2 \sqrt{s}} \label{eq:p3a} \,,
\end{align}
and $E_{3}^{\pm}(s) = \sqrt{m_{3}^{2} + p_{3}^{\pm}(s)^{2}}$ are functions of $s$, which are obtained as follows.
First, for fixed $s$, we substitute masses and momenta into Eq.~\eqref{eq:def-s}, and solve the resultant equation for $p_{3}$:
\begin{align}
  \label{eq:scat-epm}
  s = m_{3}^{2} + m_{\axino}^{2} + 2 E_{\axino} \sqrt{m_{3}^{2} + p_{3}^{2}} - 2 p_{3} p_{\axino} \cos\theta \,,
\end{align}
where $\theta$ is the angle between the three-momenta of the axino and particle 3.
Then, we vary $\cos\theta$ in the obtained solution of $p_{3}$, and find the maximum (minimum) as $p_{3}^{+}$ $(p_{3}^{-})$.

\subsubsection{3-body decay}
\label{sec:3-body-decay}

For 3-body decay, $1 \to \axino + 2 + 3$, the collision term is put into
\begin{align}
  \label{eq:col-3body}
  \frac{g_{\axino}}{E_{\axino}} C_{1 \to \axino + 2 + 3}(t, p_{\axino})
  &= \pm\frac{T}{512 \pi^{3} p_{\axino} E_{\axino}} \int dm_{2 3}^{2} \frac{1}{\sqrt{m_{2 3}^{2}} {\tilde p}_{1 \axino}} \ln{\left( \frac{1 \pm e^{- E_{1}^{-}(m_{2 3}^{2}) / T}}{1 \pm e^{- E_{1}^{+}(m_{2 3}^{2}) / T}} \right)} \int dm_{2 \axino}^{2} \sum_{\rm spin} |{\cal M}_{1 \to \axino + 2 + 3}|^{2} \,,
\end{align}
where $+$ ($-$) sign is taken when particle 1 is a fermion (boson).
The kinematic variables are defined by
\begin{align}
  m_{2 3}^{2} &= (p_{2} + p_{3})^{2} = (p_{1} - p_{\axino})^{2} \,, \label{eq:def-m23} \\
  m_{2 \axino}^{2} &= (p_{2} + p_{\axino})^{2} = (p_{1} - p_{3})^{2} \,, \\
  m_{3 \axino}^{2} &= (p_{3} + p_{a})^{2} = (p_{1} - p_{2})^{2} = m_{1}^{2} + m_{2}^{2} + m_{3}^{2} + m_{\axino}^{2} - m_{2 3}^{2} - m_{2 \axino}^{2} \,, \label{eq:def-m3a} \\
  {\tilde p}_{1 \axino} &= \frac{\sqrt{(m_{1} + m_{\axino})^{2} - m_{2 3}^{2}} \sqrt{(m_{1} - m_{\axino})^{2} - m_{2 3}^{2}}}{2 \sqrt{m_{2 3}^{2}}} \,,
\end{align}
and $E_{1}^{\pm}(m_{2 3}^{2}) = \sqrt{m_{1}^{2} + p_{1}^{\pm}(m_{2 3}^{2})^{2}}$ are functions of $m_{2 3}^{2}$, which are obtained as follows.
First, for fixed $m_{2 3}^{2}$, we substitute masses and momenta into Eq.~\eqref{eq:def-m23}, and solve the resultant equation for $p_{1}$:
\begin{align}
  \label{eq:scat-epm}
  m_{2 3}^{2} = m_{1}^{2} + m_{\axino}^{2} - 2 E_{\axino} \sqrt{m_{1}^{2} + p_{1}^{2}} + 2 p_{1} p_{\axino} \cos\theta \,,
\end{align}
where $\theta$ is the angle between the three-momenta of the axino and particle 1.
Then, we vary $\cos\theta$ in the obtained solution of $p_{1}$, and find the maximum (minimum) as $p_{1}^{+}$ $(p_{1}^{-})$.

\subsection{Phase space distributions from respective processes}
\label{sec:dist-process}

\begin{figure}
  \centering
  \begin{minipage}{0.25\linewidth}
   \begin{tikzpicture}
    \begin{feynman}
      \vertex at (0,1.5) (h) {\(H\)};
      \vertex at (1.5,1.5) (v);
      \vertex at (2.8,0.2) (tildeH){\({\widetilde H}\)};
      \vertex at (2.8,2.8) (a) {\(\axino\)};
      \diagram{(tildeH) -- (v) -- (a),
      (v) --[scalar] (h)};
    \end{feynman}
   \end{tikzpicture}
  \end{minipage}
  \begin{minipage}{0.25\linewidth}
   \begin{tikzpicture}
    \begin{feynman}
      \vertex at (0,0.8) (t) {\(t_{R}\)};
      \vertex at (0,2.2) (tildeH){\({\widetilde H}\)};
      \vertex at (3,2.2) (a) {\(\axino\)};
      \vertex at (3,0.8) (Q){\(Q_{L}\)};
      \vertex at (1.5,0.8) (v1);
      \vertex at (1.5,2.2) (v2);
      \diagram{(t) -- (v1) -- (Q),
        (tildeH) -- (v2) --(a),
        (v1) --[scalar, edge label=\(H\)] (v2)
      };
    \end{feynman}
   \end{tikzpicture}
  \end{minipage}
  \begin{minipage}{0.25\linewidth}
   \begin{tikzpicture}
    \begin{feynman}
      \vertex at (0,1.5) (w) {\({\widetilde W}\)};
      \vertex at (1.2,1.5) (v1);
      \vertex at (2.6,2.5) (h1) {\(H\)};
      \vertex at (1.9,1) (v2);
      \vertex at (3,1.6) (a){\(\axino\)};
      \vertex at (3,0.3) (h2) {\(H^{*}\)};
      \diagram{(w) -- (v1) --[scalar] (h1),
        (v1) --[edge label' = \({\widetilde H}\)] (v2) --(a),
        (v2) --[scalar] (h2)
      };
    \end{feynman}
   \end{tikzpicture}
  \end{minipage}
  \caption{Feynman diagrams for 2-body decay, $s$- or $t$-channel scattering, and 3-body decay.}
  \label{fig:diagram}
\end{figure}

Now we focus on specific examples of axino freeze-in processes, and show that the different processes result in different axino phase space distributions.
In this subsection, the following decay and scattering processes are considered.
The corresponding Feynman diagrams are shown in Fig.~\ref{fig:diagram}, and the collision terms are summarized in appendix~\ref{sec:deta-coll-term}.
\begin{figure}
  \centering
  \begin{minipage}{0.5\linewidth}
      \includegraphics[width=1.0\textwidth]{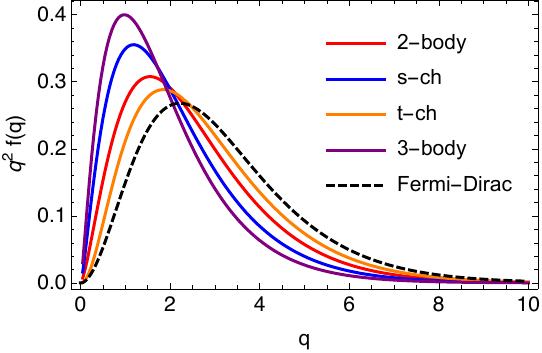}
    \end{minipage}%
    \begin{minipage}{0.5\linewidth}
      \includegraphics[width=1.0\textwidth]{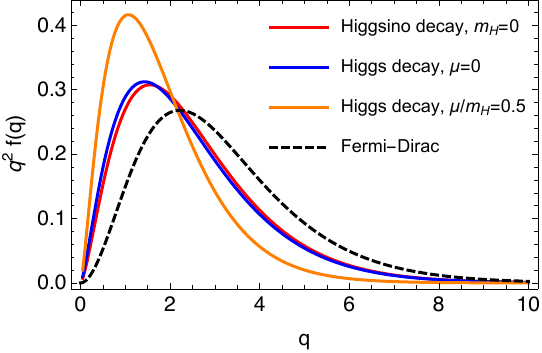}
  \end{minipage}
  \caption{Axino phase space distributions from different production processes (solid) and the Fermi-Dirac distribution (dashed) as a function of the comoving momentum, $q = p_{\axino} / T_{\axino}$.
    The phase space distribution is normalized to give $\int dq \, q^{2} f(q) = 1$.
    The left panel shows the phase space distributions from Higgsino 2-body decay, $s$-channel scattering, $t$-channel scattering, and wino 3-body decay.
    The right panel compares the phase space distributions from Higgsino decay into the axino and massless $H_{L}$, Higgs decay into the axino and massless Higgsino, and Higgs decay into the axino and massive Higgsino.}
  \label{fig:dist-process}
\end{figure}
\begin{itemize}
\item
  2-body decay of the Higgsino (left panel of Fig.~\ref{fig:diagram}, appendices~\ref{sec:2-body-decay-app} and \ref{sec:higgsino}): ${\widetilde H} \to \axino + H_{L}$.
  We assume that $m_{H_{L}} / \mu \ll 1$.

\item
  2-body decay of the lighter Higgs (left panel of Fig.~\ref{fig:diagram}, appendices~\ref{sec:higgs-2-body} and \ref{sec:higgses}): $H_{L} \to \axino + {\widetilde H}$.
  We consider the cases with $\mu / m_{H_{L}} \ll 1$ and $\mu / m_{H_{L}} = 0.5$.
  
\item
  Scattering of the Higgsino via the $s$-channel exchange of the lighter Higgs (middle panel of Fig.~\ref{fig:diagram}, appendices~\ref{sec:s-channel-scattering} and \ref{sec:higgsino}): $Q_{L}^{c} + t_{R} \to \axino + {\widetilde H}$.
  Here, $t_{R}$ is the right-handed top, $Q_{L}$ is the third-generation left-handed quark doublet, and they are taken as massless particles.
  We also assume that $H_{L}$ is massless.

\item
  Scattering of the Higgsino via the $t$-channel exchange of the lighter Higgs (middle panel of Fig.~\ref{fig:diagram}, appendices~\ref{sec:t-channel-scattering} and \ref{sec:higgsino})): ${\widetilde H} + Q_{L} \to \axino + t_{R}$ and ${\widetilde H} + t_{R}^{c} \to \axino + Q_{L}^{c}$.
  The particle content is the same as for $s$-channel scattering, but to avoid infrared divergence, we take account of the thermal mass of intermediate $H_{L}$ as $m_{H_{L}}(T) \simeq Y_{t} T / 2$, where $Y_{t} = y_{t} \cos \alpha / \sin \beta$ ($y_{t} \simeq 1$ is the top Yukawa coupling of the SM Higgs, and $\cos \alpha \simeq \sin \beta$ is the mixing between the Higgses).

\item
  3-body decay of the wino via the virtual Higgsino (right panel of Fig.~\ref{fig:diagram}, appendices~\ref{sec:3-body-decay-app} and \ref{sec:wino}): ${\widetilde W} \to \axino + H_{L} + H_{L}^{*}$.
  In this case, we assume that the Higgsino is heavier than the wino, and $H_{L}$ is massless.
\end{itemize}

Freeze-in production becomes efficient when the temperature drops to the threshold scale: $T_{\rm th} \simeq \mu$ for Higgsino decay and scattering, $T_{\rm th} \simeq m_{H_{L}}$ for decay of the lighter Higgs, and $T_{\rm th} \simeq M_{2}$ for wino decay.
Since freeze-out of heavy particles such as the Higgsino, Higgs, and wino occurs after axino freeze-in, we assume that their phase space distributions are thermal as well as those of the other SM particles.
It is convenient to define the axino temperature $T_{\axino}$ as in Eq.~\eqref{eq:nonthermal-temp} with the decoupling temperature being the threshold scale ($T_{\rm th} = \mu$, $m_{H_{L}}$, or $M_{2}$).
Since the axino temperature and momentum are simply redshifted, the comoving momentum, $q = p_{\axino} / T_{\axino}$, is independent of time.
Therefore, the phase space distribution for the comoving momentum $f_{\axino}(q)$ is constant after freeze-in. 

In Fig.~\ref{fig:dist-process}, we show the resultant axino phase space distributions $q^{2} f_{\axino}(q)$.
Each phase space distribution is normalized to give $\int dq \, q^{2} f_{\axino}(q) = 1$.
As seen in the left panel of Fig.~\ref{fig:dist-process}, axino phase space distributions differ from the Fermi-Dirac one (dashed).
We can see that the phase space distribution from 3-body decay (purple solid) is colder than the others -- higher population at low momenta -- because the typical energy given to an axino is at maximum $1/3$ of the decaying particle mass.
Therefore, we naively expect that 3-body decay is favored to relax the tension of $7 \keV$ axino DM with the Ly-$\alpha$ forest constraints.
As we will discuss in the next subsection, on the other hand, freeze-in 3-body decay cannot be a dominant process when we consider realistic models and take all the processes into account together.

In the right panel of Fig.~\ref{fig:dist-process}, we show the effects of spin of decaying particle and a mass spectrum on the phase space distributions from 2-body decay.
The difference between fermion decay (Higgsino, red solid) and boson decay (Higgs, blue solid) is small.
On the other hand, 2-body decay into a massive particle (yellow solid) produces much colder axinos than 2-body decay into a massless particle.
This is because the Higgsino mass is $1/2$ of the mass of the decaying Higgs, and the typical energy given to an axino is at maximum $3/8$ of the Higgs mass.
Therefore, it is expected that the mass degeneracy resolves the tension with the Ly-$\alpha$ forest constraints~\cite{Heeck:2017xbu}.
In realistic axino DM models, however, it is not possible due to the scattering contributions, which will be discussed in more detail at the end of Sec.~\ref{sec:mps}.

\begin{table}
  \centering
  \begin{tabular}{lcccc}\hline
    & 2-body decay & $s$-channel & $t$-channel & 3-body decay  \\ \hline
    ${\tilde \sigma}$ & 2.96 & 2.59 & 3.27 & 2.29 \\
    $a$ & 1.57 & 1.28 & 1.97 & 1.15 \\
    $b$ & 1.02 & 1.06 & 1.06 & 1.15 \\  \hline
  \end{tabular}
  \caption{${\tilde \sigma}$ calculated by the axino phase space distribution from each production process, and the fitting parameters $a$ and $b$, where the fitting function is given by $q^{2} f(q) \propto q^{a}e^{-bq}$.}
  \label{tab:sigma-fit}
\end{table}

For each phase space distribution in the left panel of Fig.~\ref{fig:dist-process}, we calculate ${\tilde \sigma}$ (see Eq.~\eqref{eq:sec_moment}) and summarize it in Table~\ref{tab:sigma-fit}.
Note that we fix the model parameters such as the Higgsino mass, although the phase space distribution and thus ${\tilde \sigma}$ depends on them as shown in the right panel of Fig.~\ref{fig:dist-process}.
In the calculation of matter power spectra in Sec.~\ref{sec:mps}, we use the fitting functions of the resultant phase space distributions.
They are parametrized as $q^{2} f(q) \propto q^{a} e^{- b q}$, where $a$ and $b$ represent the power law for low momenta and exponential (Boltzmann) suppression for high momenta, respectively.
The fitting parameters are summarized in Table~\ref{tab:sigma-fit}.
The fitting function for 2-body decay, $q^{2} f(q) \propto q^{1.57}\exp(- 1.02 q)$, agrees with the result of Ref.~\cite{Boyanovsky:2008nc}, $q^{2} f(q) \propto q^{- 1.5}\exp(- q)$.

\subsection{Phase space distributions in realistic axino DM models}
\label{sec:distr-realst-axino}

In a realistic axino DM model, the phase space distribution becomes a superposition of those from respective production channels with appropriate weights, and it also depends on the reheating temperature.
For a realistic analysis, we consider the following two benchmark points of the SUSY spectrum: one (BM1) is the case with the Higgsino being the next-to-lightest supersymmetric particle (NLSP), while the other (BM2) is the case with the wino being the NLSP.
In the BM1 case, we set $\mu = 500 \GeV$, $M_{2} = 10 \TeV$, $m_{A} = 10 \TeV$, $m_{{\widetilde Q}_{3}} = m_{{\tilde t}^{c}} = 6.5 \TeV$, and 
the masses of the other SUSY particles to $10 \TeV$.
In the BM2 case, we set $\mu = 10 \TeV$, $M_{2} = 500 \GeV$, $m_{A} = 20 \TeV$, and the masses of the other SUSY particles to $10 \TeV$.
In both cases, we take the decoupling limit, and set all $A$-terms to zero and $\tan\beta = 20$.
These spectra are summarized in Table~\ref{tab:spectra}.
\begin{table}
  \centering
  \begin{tabular}{cccc}\hline
   &   & BM1 & BM2 \\ \hline
    ~Higgs VEV ratio~ & ~$\tan\beta$~ & ~20~ & ~20~ \\
    ~$\mu$-term~ & ~$\mu$~ & ~$500 \GeV$~ & ~$10 \TeV$~ \\
    ~wino mass~ & ~$M_{2}$~ & ~$10 \TeV$~ & ~$500 \GeV$~ \\
    ~$CP$-odd Higgs mass~ & ~$m_{A}$~ & ~$10 \TeV$~ & ~$20 \TeV$~ \\
    ~stop masses~ & ~$m_{{\widetilde Q}_{3}} = m_{{\tilde t}^{c}}$~ & ~$6.5 \TeV$~ & ~$10 \TeV$~ \\
    ~SM-like Higgs mass~ &~ $m_{h}^{\rm SM-like}$~ & ~$125 \GeV$~ & ~$126 \GeV$~ \\
      \hline
    ~$H_{u}$ soft mass~&~$m_{H_{u}}^2 (Q = m_{{\tilde t}^{c}})$~ & ~$(956 \GeV)^2$~ & ~$-(9.86 \TeV)^2$~  \\
    ~$H_{d}$ soft mass~&~$m_{H_{d}}^2 (Q = m_{{\tilde t}^{c}})$~ & ~$(9.94 \TeV)^2$~ & ~$(17.3 \TeV)^2$~ \\
    \hline
  \end{tabular}
  \caption{MSSM parameters in the BM1 and BM2 cases.
  The SM-like Higgs mass and soft masses at $Q = m_{{\tilde t}^{c}}$ are calculated by \texttt{SUSY-HIT v1.5a}~\cite{Djouadi:2006bz}. 
  The masses of the other SUSY particles are taken to be $10 \TeV$.}
  \label{tab:spectra}
\end{table}

\begin{figure}
  \centering
  \begin{minipage}{0.5\linewidth}
   \includegraphics[width=1.0\textwidth]{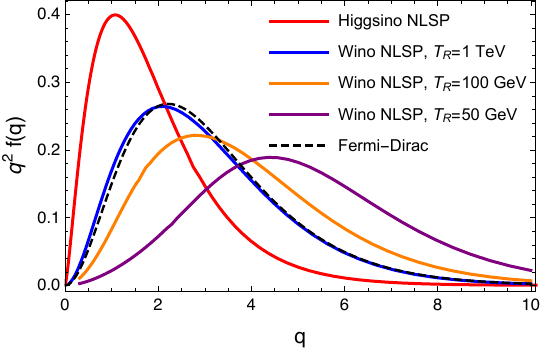}  
 \end{minipage}%
 \begin{minipage}{0.5\linewidth}
    \includegraphics[width=1.0\textwidth]{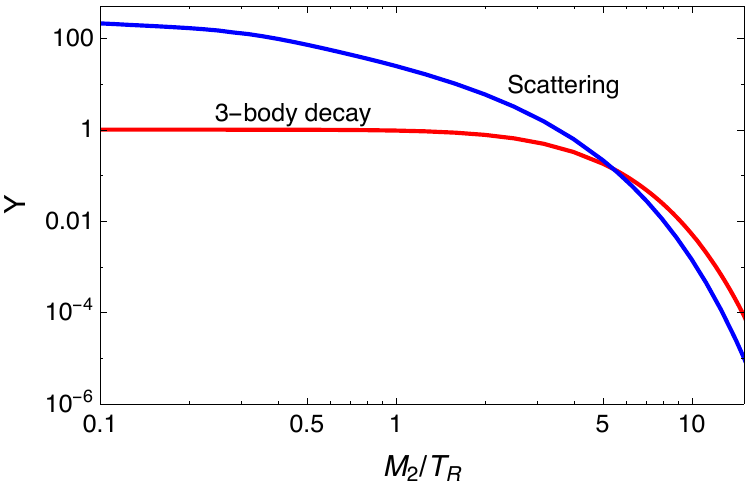} 
  \end{minipage}
  \caption{
    Left: Axino phase space distributions in realistic models.
    Normalization is the same as in Fig.~\ref{fig:dist-process}.
    For comparison the Fermi-Dirac distribution (dashed) is also plotted.
    Right: Axino yields from wino 3-body decay (red) and scattering (blue), ${\widetilde W} + H_{L}^{(*)} \to \axino + H_{L}^{(*)}$ and $H_{L} + H_{L}^{*} \to \axino + {\widetilde W}$.
  }
  \label{fig:dist-real}
\end{figure}

\subsubsection{Higgsino NLSP (BM1)}
\label{sec:higgsino-2-body}

In the BM1 case, we assume that the reheating temperature is higher than the Higgs mass.
Axino freeze-in occurs via Higgs 2-body decay, Higgsino $s$-channel scattering, and $t$-channel scattering.
Among them, 2-body decay of $H_{L}$ is dominant due to the largest phase space factor, as we can see by comparing Eqs.~\eqref{eq:col-2body} and \eqref{eq:col-scat}.
The left panel of Fig.~\ref{fig:dist-real} shows that the resultant phase space distribution (red solid) is similar to that of Higgs 2-body decay into the axino and massive Higgsino (yellow solid) in the right panel of Fig.~\ref{fig:dist-process}.

\subsubsection{Wino NLSP with a high reheating temperature (BM2 with $M_{2} < T_{R} < \mu$)}
\label{sec:wino-scatt-domin}

In the BM2 case, the Higgsino is much heavier than the wino.
We also assume that $M_{2} < T_{R} < \mu$.
The wino diagram in the right panel of Fig.~\ref{fig:diagram} implies that we have to consider wino scattering as well as its 3-body decay, and have to figure out which process is dominant.%
\footnote{As shown in Table~\ref{tab:spectra}, $m_{H_{u}}^2$ at the stop mass scale is negative. 
However, the mass term of $H_{u}$ is $m_{H_{u}}^2 + |\mu|^2$, and it is positive at that scale.
At the lower temperature where wino decay and scattering become important, $T \sim M_{2}$, we assume that the lighter Higgs can be regarded as massless. 
If we turn on the mass of the lighter Higgs, {\it e.g.}, $m_{H_{L}} = 125 \GeV$, the resultant phase space distribution of axinos becomes slightly colder,
but it does not alter our conclusion.}
The right panel of Fig.~\ref{fig:dist-real} compares the axino yields from wino scattering (blue) and 3-body decay (red).
Each yield is normalized by the yield from 3-body decay with a sufficiently high reheating temperature.
We can see that for $M_{2} < 5 T_{R}$, wino scattering such as ${\widetilde W} + H_{L} \to \axino + H_{L}$ dominates axino production.
Since the mass of the intermediate Higgsino is much larger than the typical energy transfer, scattering is effectively described by a dimension-5 operator, and therefore axinos are dominantly produced at the time of reheating, not at $T \simeq M_{2}$.

As seen in the left panel of Fig.~\ref{fig:dist-real}, the BM2 scenario with $T_{R} = 1 \TeV$ ($M_{2} / T_{R} = 0.5$) results in a phase space distribution (blue solid) close to the Fermi-Dirac one (dashed).
This can be understood as follows.
Since axinos are produced mainly at $T \sim T_{R}$ ($M_{2} < T_{R} < \mu$), the squared matrix element is roughly given by $\sum_{\rm spin} |{\cal M}|^{2} \sim (g_{2} \mu / v_{\rm PQ})^{2} \times s / \mu^{2} \sim g_{2}^{2} s / v_{\rm PQ}^{2}$.
Then, Eq.~\eqref{eq:col-scat} is approximately given by
\begin{equation}
  \label{eq:wino-scat-app}
  \frac{g_{\axino}}{E_{\axino}}C_{1 + 2 \to \axino + 3}(t, p_{\axino})
  \sim
  \frac{T}{p_{\axino}^{2}} e^{-p_{\axino} / T} \int ds g_{2}^{2} \frac{s}{v_{\rm PQ}^{2}}
  \sim g_{2}^{2} \frac{T^{3}}{v_{\rm PQ}^{2}} e^{- p_{\axino} / T} \,,
\end{equation}
where we use $s \sim p_{\axino} T$ as the thermal average.
Substituting it into Eq.~\eqref{eq:integ-boltz}, we obtain the Boltzmann distribution for axino, $f_{\axino} \sim e^{- p_{\axino} / T}$.
Note that as long as $M_{2} < T_{R} < \mu$, the shape of the resultant phase space distribution is independent of $T_{R}$.

\subsubsection{Wino NLSP with a low reheating temperature (BM2 with $M_{2} > T_{R}$)}
\label{sec:wino-3-body}

The NLSP is the wino as in the previous scenario, namely in the BM2 case.
We, however, assume that the reheating temperature is below the wino mass, $T_{R} = M_{2} / 5$.
With such a low reheating temperature, the right panel of Fig.~\ref{fig:dist-real} shows that the axino yield from wino 3-body decay (red) is comparable to that from scattering (blue).
Wino 3-body decay produces 46\% of the total axino abundance.
The resultant phase space distribution (yellow solid) in the left panel of Fig.~\ref{fig:dist-real} is, however, different from that of freeze-in 3-body decay shown in Fig.~\ref{fig:dist-process}.
This is because with $T_{R} = 100 \GeV$ ($M_{2} / T_{R} = 5$), axino production occurs at $T \sim T_{R} = M_{2} / 5$, and the typical energy given to an axino, which is roughly $M_{2} / 3$, is larger than the temperature.
As a result, the axino phase space distribution becomes hotter than the Fermi-Dirac one (dashed).
In order to elucidate this point, the BM2 scenario with $T_{R} = 50 \GeV$ ($M_{2} / T_{R} = 10$, purple solid) is also shown in Fig.~\ref{fig:dist-real}, where the effects of the electroweak phase transition is ignored for simplicity.
The resultant phase space distribution is much hotter, and unlike the previous scenario ($M_{2} < T_{R} < \mu$), the shape of the distribution depends on $T_{R}$ in this case.

As we will see in the next section, a colder phase space distribution relaxes the tension of $7 \keV$ axino DM with Ly-$\alpha$ forest data.
Figure~\ref{fig:dist-process} apparently shows that freeze-in 3-body decay (purple solid) gives the coldest phase space distribution, and we may expect that the wino NLSP in a decoupled Higgsino scenario is favored as a realistic $7 \keV$ axino model.
In general, however, 2-to-2 scattering is induced by the same diagram as 3-body decay through crossing symmetry.
Although the phase space factor is similar, the scattering rate increases with the temperature, while the 3-body decay rate does not.
Therefore, unless the reheating temperature is as low as $T_{R} \lesssim M_{2} / 5$, 2-to-2 scattering dominates axino production, and the resultant phase space distribution is almost the same as the Fermi-Dirac one.
Even when the reheating temperature is low so that 3-body decay dominates axino production, axinos are dominantly produced at $T \sim T_{R}$, not $T \sim M_{2}$, which results in the phase space distribution hotter than the Fermi-Dirac one.
Therefore, whatever the reheating temperature is, we cannot obtain a colder phase space distribution than the Fermi-Dirac one.
This is not special to wino 3-body decay into the axino.
One always encounters such a difficulty for 3-body decay through heavy intermediate particles, and cannot turn to 3-body decay for a cold phase space distribution.

\section{Linear matter power spectrum}
\label{sec:mps}

In this section, we relate the axino phase space distribution to the observed matter distribution of the Universe, especially, Ly-$\alpha$ forest data.
We solve the evolution equation of the cosmological perturbations by incorporating DM phase space distributions in \texttt{CLASS}~\cite{Blas:2011rf, Lesgourgues:2011rh} with the cosmological parameters from ``Planck 2015 TT, TE, EE+lowP" in Ref.~\cite{Ade:2015xua}.
The resultant linear matter power spectra are cross-checked with \texttt{CAMB}~\cite{Lewis:1999bs} by suitably incorporating the covariant multipole perturbation method~\cite{Ma:1995ey, Lewis:2002nc}.
By implementing the Fermi-Dirac distribution, we obtain the matter power spectra with $m_{\rm WDM} = 2.0$, $3.3$, $4.09$, and $5.3 \keV$, which represent the Ly-$\alpha$ forest constraints.
To implement the axino phase space distributions from respective production processes, we use the fitting functions of $q^{2} f_{\axino}(q) \propto q^{a} e^{- b q}$ with the fitting parameters in Table~\ref{tab:sigma-fit}.
For realistic axino DM models, we also fit the phase space distributions in the left panel of Fig.~\ref{fig:dist-real}, whose fitting parameters are summarized in Table~\ref{tab:fit-bm} of appendix~\ref{sec:fitt-fuct-benchm}.
Throughout the analyses, the axino mass is fixed at $7 \keV$. 

We follow the analysis suggested in Ref.~\cite{Konig:2016dzg} when constraining $7 \keV$ freeze-in axino DM by Ly-$\alpha$ forest data.
Given a matter power spectrum, $P(k)$ ($k$ is the wavenumber), we define a squared transfer function by
\begin{align}
  \label{eq:transfer}
  {\cal T}^{2}(k) = \frac{P(k)}{P_{\rm CDM}(k)} \,,
\end{align}
where $P_{\rm CDM}(k)$ is the CDM matter power spectrum.
We compare the squared transfer function of axino DM, ${\cal T}^{2}_{\axino}(k)$, to that of the conventional WDM, ${\cal T}^{2}_{\rm WDM} (k)$.
If ${\cal T}^{2}_{\axino} (k) < {\cal T}^{2}_{\rm WDM} (k)$ is met for any $k$, the axino DM model is regarded as being excluded.
This naive determination is, however, sometimes not applicable, because the slopes of ${\cal T}^{2}(k)$ above the cutoff scale are different between thermal (conventional WDM) and non-thermal (axino DM) phase space distributions, and ${\cal T}^{2}_{\axino} (k) < {\cal T}^{2}_{\rm WDM} (k)$ holds only for some range of $k$.
In such a case, we first determine the half-mode $k_{1/2}$ by ${\cal T}^{2}_{\axino}(k_{1/2}) = 1 / 2$.
Then, if ${\cal T}^{2}_{\axino}(k) < {\cal T}^{2}_{\rm WDM}(k)$ is met for all $k < k_{1/2}$, we regard the axino DM model as being excluded.

\begin{figure}
  \centering
  \begin{minipage}{0.5\linewidth}
      \includegraphics[width=1.0\textwidth]{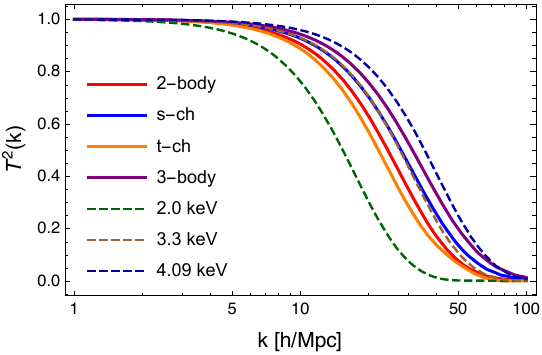}
  \end{minipage}%
  \begin{minipage}{0.5\linewidth}
    \includegraphics[width=1.0\textwidth]{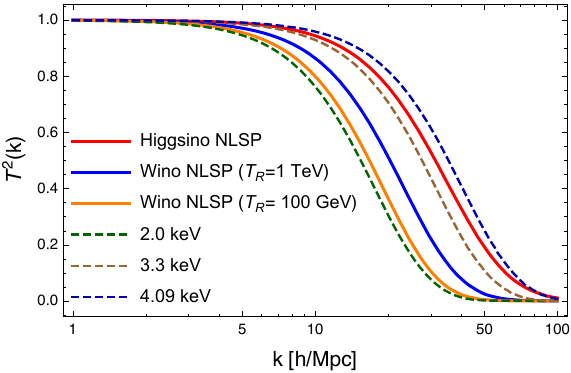}
  \end{minipage}
  \caption{Squared transfer functions for respective production processes (left, solid) and in realistic axino DM models (right, solid).
                The conventional WDM models with $m_{\rm WDM} = 2.0$, $3.3$, and $4.09 \keV$ are shown for comparison (dashed).}
  \label{fig:T2k}
\end{figure}

Figure~\ref{fig:T2k} shows ${\cal T}^{2}(k)$ as a function of $k \, [h / \rm Mpc]$.
In the left panel, we compare the squared transfer functions for respective production processes discussed in Sec.~\ref{sec:dist-process}.
Each ${\cal T}^{2}(k)$ corresponds to each phase space distribution in the left panel of Fig.~\ref{fig:dist-process}.
For comparison, we show $2.0$ (green dashed), $3.3$ (brown dashed), and $4.09 \keV$ WDM (blue dashed) as Ly-$\alpha$ forest constraints.
The $7 \keV$ axino DM relics from Higgsino 2-body decay (red solid) and $t$-channel scattering (yellow solid) are inconsistent with the Ly-$\alpha$ forest constraint of $m_{\rm WDM} \gtrsim 3.3 \keV$, since the cutoff scale is clearly smaller.
On the other hand, that from Higgsino $s$-channel scattering (blue solid) seems as warm as $3.3 \keV$ WDM, and that from wino 3-body decay (purple solid) is consistent with the constraint.

Interestingly, we can infer these results by the discussion of warmness introduced in Sec.~\ref{sec:intro_wdm}.
Using the relation in Eq.~\eqref{eq:mass-rel} and ${\tilde \sigma}$ in Table~\ref{tab:sigma-fit}, we obtain $2.5 \keV ({\tilde \sigma} / 3.6)^{-3/4} = 2.9$, $3.2$, $2.7$, and $3.6 \keV$, respectively, for the $7 \keV$ axino DM relics from 2-body decay, $s$-channel scattering, $t$-channel scattering, and 3-body decay.
Therefore if we take $m_{\rm WDM} \gtrsim 3.3 \keV$ as a Ly-$\alpha$ forest constraint, we expect that the $7 \keV$ axino DM relics from 2-body decay and $t$-channel scattering are inconsistent, that from $s$-channel scattering is comparable, and that from 3-body decay is consistent with the Ly-$\alpha$ forest constraint.
The expectation agrees with the result that we obtain by computing and comparing the squared transfer functions.
This analytic method through Eq.~\eqref{eq:mass-rel} is far simpler and provides the direct correspondence between the phase space distributions and the mass bounds.

We also calculate the linear matter power spectra in the realistic axino DM models studied in Sec.~\ref{sec:distr-realst-axino}, which are shown in the right panel of Fig.~\ref{fig:T2k}.
In the BM1 (Higgsino NLSP) scenario, axino DM production is dominated by 2-body decay of the lighter Higgs into the axino and massive Higgsino (red solid).
In the BM2 (wino NLSP) scenario, depending on the reheating temperature, axinos are dominantly produced by wino 3-body decay ($T_{R} = M_{2} / 5$, yellow solid) or scattering ($T_{R} > M_{2}$, blue solid).
Both the matter power spectra in the BM2 scenarios are hotter than that in the BM1 scenario.
In Eq.~\eqref{eq:mass-rel}, we find that $2.5 \keV ({\tilde \sigma} / 3.6)^{-3/4} = 3.6$, $2.5$, and $2.1 \keV$ in the BM1 scenario and BM2 scenarios with $M_{2} / T_{R} = 0.5$ and $M_{2} / T_{R} = 5$, respectively.
All the scenarios have tension with the Ly-$\alpha$ forest constraint of $m_{\rm WDM} \gtrsim 4.09 \keV$ (blue dashed).
Furthermore, both the BM2 scenarios are disfavored even by the weaker constraint of $m_{\rm WDM} \gtrsim 3.3 \keV$ (brown dashed), while the BM1 scenario is consistent with it.

\begin{figure}
  \centering
  \includegraphics[width=0.7\textwidth]{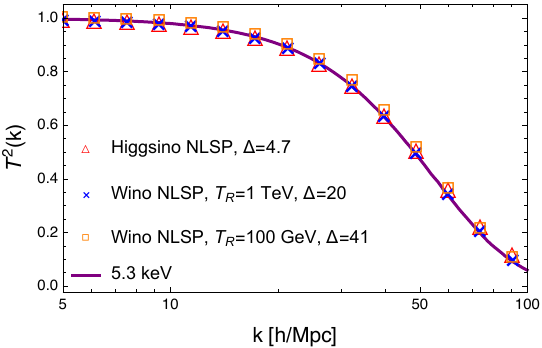}
  \caption{Squared transfer functions in realistic axino DM models with entropy production $\Delta$ (colored markers).
  The conventional WDM model with $m_{\rm WDM} = 5.3 \keV$ is shown for comparison (purple solid line).}
  \label{fig:T2k-entropy}
\end{figure}

However, in the SUSY axion model, inherent entropy production from saxion decay mitigates the tension of $7 \keV$ axino DM with Ly-$\alpha$ forest constraints.
Entropy production changes the axino DM temperature (see Eq.~\eqref{eq:nonthermal-temp}) after saxion decay, $T < T_{D}^{s}$, as
\begin{align}
  \label{eq:axino-temp-ent-pro}
  T_{\axino} = \left( \frac{g_{*}(T)}{\Delta \, g_{*}(T_{\rm th})} \right)^{1/3} T \,,
\end{align}
where $\Delta$ is the entropy dilution factor given by Eq.~\eqref{eq:dil-fac}.
Here we assume that the saxion dominates the energy density of the Universe after axino decoupling, $T_{e}^{s} < T_{\rm th}$.
In such a case, saxion domination and subsequent decay do not distort the axino phase space distribution, and thus we can use $f(q)$'s calculated in Sec.~\ref{sec:mom-dist-axino}.
With this correction from entropy production, Eq.~\eqref{eq:mass-rel} turns into
\begin{align}
  \label{eq:mass-rel-ent-pro}
  m = 7 \keV \left( \frac{m_{\rm WDM}}{2.5 \keV \Delta^{1/4} ({\tilde \sigma} / 3.6)^{-3/4}} \right)^{4/3} \,.
\end{align}
From this, we can infer the minimum value of the entropy dilution factor required to avoid the Ly-$\alpha$ forest constraint: $\Delta > \left[ m_{\rm WDM} / (2.5 \keV ({\tilde \sigma} / 3.6)^{-3/4}) \right]^{4}$.
For instance, to take account of the Ly-$\alpha$ forest constraint of $m_{\rm WDM} \gtrsim 5.3 \keV$, $\Delta$ should be larger than $4.7$, $20$, and $41$, respectively, in the BM1 scenario and BM2 scenarios with $M_{2} / T_{R} = 0.5$ and $M_{2} / T_{R} = 5$.
In Fig.~\ref{fig:T2k-entropy}, we confirm that the resultant matter power spectra in realistic axino DM models with those values of $\Delta$ are comparable with that of the conventional WDM model with $m_{\rm WDM} = 5.3 \keV$.

Such entropy production from saxion decay also changes an axino overabundance.
The total axino yields from the phase space distributions calculated in Sec.~\ref{sec:distr-realst-axino} are given by
\begin{eqnarray}
Y_{\axino} \simeq \left( \frac{10^{10} \GeV}{v_{\rm PQ}} \right)^{2} \left\{
\begin{matrix}
& 1.8 \times 10^{-3} & \qquad \text{(BM1)} \,, \\
& 2.2 & \qquad \text{(BM2 with $T_{R} = 1 \TeV$)} \,, \\
& 1.2 \times 10^{-2} & \qquad \text{(BM2 with $T_{R} = 100 \GeV$)} \,,
\end{matrix} \right.
\end{eqnarray}
without entropy production.
From the observed density parameter, $\Omega_{\axino} h^{2} = 2.0 \times 10^{3} \, Y_{\axino} \left({m_{\axino}}/{7 \keV}\right) = 0.12$, the yield of $7 \keV$ axino DM has to be $Y_{\axino} = 6.1 \times10^{-5}$.
Therefore, the dilution factors are related with the PQ-breaking scale as
\begin{equation}
\Delta \simeq \left\{
\begin{matrix}
& 4.7 \left({2.5 \times 10^{10} \GeV} / {v_{\rm PQ}} \right)^{2} & \qquad \text{(BM1)} \,, \\
& 20 \left({4.2 \times 10^{11} \GeV} / {v_{\rm PQ}} \right)^{2} & \qquad \text{(BM2 with $T_{R} = 1 \TeV$)} \,, \\
& 41 \left({2.2 \times 10^{10} \GeV} / {v_{\rm PQ}} \right)^{2} & \qquad \text{(BM2 with $T_{R} = 100 \GeV$)} \,.
\end{matrix}\right.
\end{equation}
In the BM1 scenario, the dilution factor of $\Delta = 4.7$ can be easily obtained from saxion decay.
For $v_{\rm PQ} = 2.5 \times 10^{10}$\,GeV, the saxion with the mass around $110 \GeV$
dominantly decays into the $b$-quark pair via a saxion-Higgs mixing (see Ref.~\cite{Bae:2013hma} for details of saxion decay).
In such a case, the decay temperature is $T_{D}^{s} \simeq 53 \GeV$.
In the meantime, as shown in Eq.~\eqref{eq:eq_tem}, saxion domination occurs at $T^{s}_{e} \simeq 250 \GeV$ when $s_{0} = 10^{16} \GeV$.
Thus, the dilution factor is $\Delta = T^{s}_{e} / T^{s}_{D} \simeq 4.7$.
In the BM2 scenarios, on the other hand, saxion domination occurs at a very low temperature, $T^{s}_{e} \sim 2.5 \text{--} 25 \MeV$, since we consider a low reheating temperature, $T_{R} < \mu$.
Although one can consider the very light saxion whose decay temperature is of MeV order or smaller, entropy injection at such a low temperature may be disfavored by the big bang nucleosynthesis.
If the reheating temperature is larger, one can obtain larger $T^{s}_{e}$.
In such a case, the Higgsino contributions dominate axino production, and thus the basic feature becomes the same as in the BM1 scenario.

\begin{figure}
  \centering
  \includegraphics[width=0.7\textwidth]{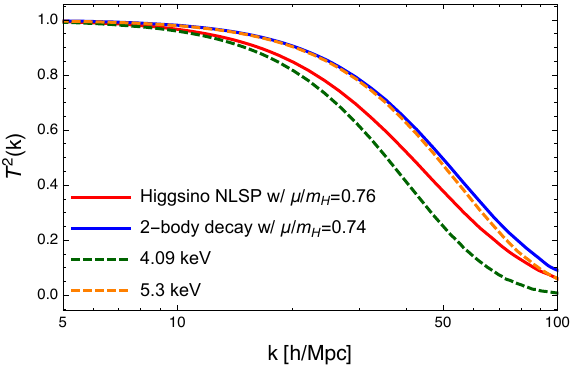}
  \caption{Squared transfer functions for Higgs 2-body decay and Higgsino scattering with $\mu / m_{H_{L}} = 0.76$ (red solid), and only Higgs 2-body decay with $\mu / m_{H_{L}} = 0.74$ (blue solid).
    The conventional WDM models with $m_{\rm WDM} = 4.09$ and $5.3 \keV$ are shown for comparison (dashed).}
  \label{fig:mass-degen}
\end{figure}

Another way to evade the Ly-$\alpha$ forest constraints may be to take a degenerate mass spectrum~\cite{Heeck:2017xbu}.
As is evident in Fig.~\ref{fig:dist-process}, the axino phase space distribution from 2-body decay becomes colder when the mass difference between decaying particle and its decay products is smaller.
Thus we expect that in the BM1 scenario, the tension with Ly-$\alpha$ may be relaxed by tuning $\mu$ and $m_{H_{L}}$.
However, the axino yield from 2-body decay decreases as the mass difference decreases, while that from Higgsino scattering is not affected so much.
Therefore, the phase space distribution for a very small mass difference is dominated by $s$-channel scattering and $t$-channel scattering, and thus it cannot be arbitrarily cold.

To demonstrate it quantitatively, we consider a similar model based on the BM1 scenario, where $\mu$ is taken as arbitrary.
We find that $\mu / m_{H_{L}} = 0.76$ provides the coldest phase space distribution in terms of ${\tilde \sigma}$, which gives $2.5 \keV({\tilde \sigma} / 3.6)^{-3/4} = 4.2 \keV$ in Eq.~\eqref{eq:mass-rel}.%
\footnote{We take account of only the lighter Higgs here. If the mass difference between the Higgses is similar to that in the BM1 scenario, the heavier Higgs provides about 10 \% correction mainly from its 2-body decay.}
We compute its linear matter power spectrum and show ${\cal T}^{2}(k)$ (red solid) in Fig.~\ref{fig:mass-degen}.
Unlike the BM1 scenario, the Ly-$\alpha$ forest constraint of $m_{\rm WDM} \gtrsim 4.09 \keV$ is evaded, although the most stringent constraint of $m_{\rm WDM} \gtrsim 5.3 \keV$ cannot be evaded.
This is because scattering dominates axino production for $\mu / m_{H_{L}} \gtrsim 0.8$ due to the large top Yukawa coupling, which is not a free parameter in the axino model.
If we could choose the coupling freely, the scattering contributions could be small enough to evade even the most stringent constraint.
To see this, we consider the case where only Higgs decay exists and Higgsino scattering is switched off.
We find that the most stringent Ly-$\alpha$ forest constraint is evaded for $\mu / m_{H_{L}} \gtrsim 0.74$.
The matter power spectrum for $\mu / m_{H_{L}} = 0.74$ (blue solid) is also shown in Fig.~\ref{fig:mass-degen}.
Note that for a small mass difference, thermal effects such as the thermal mass and width become important at least for the total abundance~\cite{Hamaguchi:2011jy}.
Studying such effects at the level of the phase space distribution is beyond the scope of this paper.

\section{Conclusions}
\label{sec:concl}

While decaying $7 \keV$ DM is one of the most promising explanations of the $3.5 \keV$ line excess, one needs to care the warmness of such DM especially in response to the Ly-$\alpha$ forest observations probing the smaller-scale matter distribution of the Universe.
Although the Ly-$\alpha$ forest constraints (even the most stringent constraint of $m_{\rm WDM} \gtrsim 5.3 \keV$) apparently allow $7 \keV$ DM, a very low DM temperature is implicitly assumed; {\it i.e.}, we need $g_{*, {\rm WDM}} \sim 7000$ (to be compared with the number of degrees of freedom before the electroweak phase transition, $g_{*} = 106.75$), which requires a large entropy dilution factor after DM decoupling, $\Delta \sim 70$.
For DM particles decoupled when $g_{*} = 106.75$, DM with the mass of $m = 7 \keV \left(m_{\rm WDM} / 2.5 \keV / ({\tilde \sigma} / 3.6)^{-3/4} \right)^{4/3}$ has the same warmness as the conventional WDM with the mass $m_{\rm WDM}$.
Thus, such $7 \keV$ DM is in tension with the Ly-$\alpha$ forest constraint of $m_{\rm WDM} \gtrsim 3.3 \keV$. 

Not only the temperature, but also the phase space distribution of DM also affects the resultant warmness of DM through its dimensionless divergence ${\tilde \sigma}$ in the above mass relation, where ${\tilde \sigma} = 3.6$ is the reference value for the Fermi-Dirac distribution of the conventional WDM.
In the previous literature (that mainly focused on sterile neutrino DM), the phase space distribution of freeze-in DM was calculated, while discussion at the level of linear matter power spectra has been scarce in number.
The phase space distribution of freeze-in DM is typically colder than the Fermi-Dirac distribution of the conventional WDM.
We have compared linear matter power spectra directly to examine if such a colder distribution mitigates the tension with Ly-$\alpha$ forest data.
Furthermore, formulas for various freeze-in processes collected in Sec.~\ref{sec:boltzmann-eq} and appendix~\ref{sec:deta-coll-term} are useful in studying other freeze-in DM models.

In this paper, specifically, we have considered freeze-in production of axino DM in the SUSY DFSZ model in order to resolve this tension.
In our model, $7 \keV$ axino DM decays into photons and neutrinos via bRPV operators.
Axinos are mainly produced by heavy particle decay and/or scattering processes of particles in the thermal bath.
Due to its apparently renormalizable couplings, axinos are dominantly produced at the temperature near the threshold scale, so it is perfectly matched with the freeze-in DM scenario.
Since the scale is of the order of the $\mu$-term, production processes take place before the electroweak phase transition.
By calculating the Boltzmann equation, we have found the following results.
\begin{itemize}
\item[1.] The phase space distributions are different depending on production processes such as 2-body decay, $s$-channel scattering, $t$-channel scattering, and 3-body decay.
When we consider the respective processes separately, the axino DM relics from 3-body decay show the coldest phase space distribution, while those from 2-body decay, $s$-channel scattering, and $t$-channel scattering show hotter ones (see Fig.~\ref{fig:dist-process}).
All cases show colder phase space distributions than the typical thermal case.

\item[2.] We have shown three realistic scenarios with two benchmark points:
the Higgsino NLSP (BM1) scenario with $T_{R} > m_{H_{L}}$ and wino NLSP (BM2) scenarios with $T_{R} > M_{2}$ and $T_{R} < M_{2}$ (see Fig.~\ref{fig:dist-real}).
In the BM1 scenario, the dominant production process is 2-body decay of the lighter Higgs into the axino and Higgsino, so the phase space distribution is colder than the Fermi-Dirac one.
In the BM2 scenario with $T_{R} > M_{2}$, however, the phase space distribution is similar to the Fermi-Dirac one. 
The reason is that production is governed by the dimension-5 operator suppressed by the large $\mu$-term, so the dominant production occurs at the highest temperature, $T = T_{R}$.
In the BM2 scenario with $T_{R} < M_{2}$, axinos are mainly produced by 3-body decay of the wino.
On the contrary to Fig.~\ref{fig:dist-process}, it shows a hotter phase space distribution than the Fermi-Dirac one, since wino decay occurs at the temperature smaller than its mass.

\item[3.] The matter power spectra are obtained from the calculated phase space distributions.
Figure~\ref{fig:T2k} shows that both the BM1 and BM2 scenarios still have tension with the constraints of $m_{\rm WDM} \gtrsim 4.09 \keV$.
In order to avoid the tension, we need dilution factors of $\Delta \simeq 4.7$, $20$, and $41$, in the BM1 scenario and BM2 scenarios with $T_{R} = 1 \TeV$ and $T_{R} = 100 \GeV$, respectively.
In the BM1 scenario, the mild dilution factor of $\Delta = 4.7$ can be obtained by saxion decay into the $b$-quark pair if $v_{\rm PQ} \simeq 2.5 \times 10^{10} \GeV$, $s_{0} \simeq 10^{16} \GeV$ and $m_{s} \simeq 110 \GeV$, which is inherent in the SUSY DFSZ axion model.
In this case, axinos from freeze-in production meet the observed DM density.
In BM2, however, it is difficult to obtain enough dilution factors without spoiling the success of the standard cosmology in the big bang nucleosysnthesis.
We have also discussed the mass degeneracy to avoid the Ly-$\alpha$ forest constraints, and found that taking $\mu / m_{H_{L}} \simeq 0.76$ in the BM1 scenario, it is possible to evade the constraint of $m_{\rm WDM} \gtrsim 4.09 \keV$, while that of $m_{\rm WDM} \gtrsim 5.3 \keV$ cannot be evaded.
For $\mu / m_{H_{L}} \gtrsim 0.76$, the scattering contribution dominates the total axino yield, and thus the phase space distribution cannot become colder for larger $\mu / m_{H_{L}}$.
\end{itemize}

We have shown how freeze-in production of DM differs from the conventional thermal WDM.
When 2-body decay dominates DM production, it produces a colder phase space distribution and thus relieve the tension with Ly-$\alpha$ forest data.
3-body decay is accompanied by the scattering processes since they have kinematic factors and couplings of the same order.
In our example with a high reheating temperature, this case shows a phase space distribution similar to the thermal one in the conventional thermal DM.
For a low reheating temperature, 3-body decay can dominate DM production, but it produces a hotter phase space distribution.
Although freeze-in DM relieves the tension with Ly-$\alpha$ in the BM1 scenario and BM2 scenario with $T_{R} > M_{2}$, mild dilution factors are still required to avoid the strongest Ly-$\alpha$ forest constraint.
While it is possible to achieve such a dilution factor from saxion decay in the BM1 scenario, it is difficult in the BM2 scenario with a low reheating temperature.

\acknowledgements

The work of KJB and AK was supported by IBS under the project code IBS-R018-D1.
SPL has received support from the Marie-Curie program and the European Research Council and Horizon 2020 Grant, contract No. 675440 (European Union).
AK and SPL would like to acknowledge the Mainz institute for Theoretical Physics (MITP) where this work was initiated.

\appendix

\section{Details of the Boltzmann equation}
\label{sec:boltzmann-equation}

In this appendix, we present further details of the Boltzmann equation used in Sec.~\ref{sec:mom-dist-axino}.
We also discuss the validity of ignoring the Bose-enhancement and Pauli-blocking, $1 \pm f^{\rm eq} \simeq 1$.

\subsection{Derivation of Eq.~\eqref{eq:col-scat}}
\label{sec:derivation-eq}

Here we derive the collision term for scattering, Eq.~\eqref{eq:col-scat}.
The collision term for 3-body decay, Eq.~\eqref{eq:col-3body}, can also be obtained in the similar way.
We begin with the following collision term for 2-to-2 scattering:
\begin{align}
  \label{eq:col-sccat}
  \frac{g_{\axino}}{E_{\axino}} C_{1 + 2 \to \axino + 3}(t, p_{\axino})
  =&
  \frac{1}{2E_{\axino}}
    \int \prod_{i = 1}^{3} \frac{d^{3} p_{i}}{(2 \pi)^{3} 2 E_{i}}
    (2 \pi)^{4} \delta^{4}(\hat{p}_{1} + \hat{p}_{2} - \hat{p}_{\axino} - \hat{p}_{3})
    \notag\\
  &\times\sum_{\rm spin} |{\cal M}_{1 + 2 \to \axino + 3}|^{2}
    f^{\rm eq}_{1} f^{\rm eq}_{2} (1 \mp f^{\rm eq}_{3}) \,.
\end{align}
The delta function is put into the form of
\begin{align}
  \label{eq:delta}
  \frac{d^{3}p_{1}}{(2 \pi)^{3} 2 E_{1}}\frac{d^{3} p_{2}}{(2 \pi)^{3} 2 E_{2}}(2 \pi)^{4}\delta^{4}(\hat{p}_{1} + \hat{p}_{2} - \hat{p}_{\axino} - \hat{p}_{3})
  =
  \frac{dt}{16 \pi \sqrt{s} p_{3 \axino}} \,.
\end{align}
The energy conservation ensures the identity,
\begin{align}
  \label{eq:dist-balance}
  f^{\rm eq}_{1} f^{\rm eq}_{2} (1 \mp f^{\rm eq}_{3})
  =
  (1 \mp f^{\rm eq}_{1})(1 \mp f^{\rm eq}_{2})f^{\rm eq}_{3} e^{- E_{\axino} / T} \,.
\end{align}
The collision term is then written as
\begin{align}
  \label{eq:col-2}
  \frac{g_{\axino}}{E_{\axino}} C_{1 + 2 \to \axino + 3}(t, p_{\axino})
  &\simeq
  \frac{1}{2 E_{\axino}}
    e^{- E_{\axino} / T} \int \frac{d^{3} p_{3}}{(2 \pi)^{3} 2 E_{3}}
    \frac{f^{\rm eq}_{3}(E_{3})}{16 \pi \sqrt{s} p_{3 \axino}}
    \int dt \sum_{\rm spin} |{\cal M}_{1 + 2 \to \axino + 3}|^{2} \,.
\end{align}
The phase space integration, $d^{3} p_{3} = 2 \pi p_{3}^{2} d\cos \theta_{3} dp_{3}$, can be transformed into the integration with respect to $s$ and $E_{3}$,
\begin{align}
  \label{eq:integ-trsfm}
  ds dE_{3} = \frac{2 p_{\axino} p_{3}^{2}}{E_{3}} d\cos \theta_{3} dp_{3} \,.
\end{align}
The collision term is written as
\begin{align}
  \label{eq:col-3}
  \frac{g_{\axino}}{E_{\axino}} C_{1 + 2 \to \axino + 3}(t, p_{\axino})
  &\simeq
    \frac{e^{- E_{\axino} / T}}{512 \pi^{3} p_{\axino} E_{\axino}}
    \int ds \int_{E_{3}^{-}(s)}^{E_{3}^{+}(s)} dE_{3}
    \frac{1}{\sqrt{s} p_{3 \axino}} f^{\rm eq}_{3}(E_{3})
    \int dt \sum_{\rm spin} |{\cal M}_{1 + 2 \to \axino + 3}|^{2}
    \notag\\
  &=
    \pm \frac{T e^{- E_{\axino} / T}}{512 \pi^{3} p_{\axino} E_{\axino}}
    \int ds 
    \frac{1}{\sqrt{s} p_{3 \axino}} 
    \ln{\left( \frac{1\pm e^{- E_{3}^{-}(s) / T}}{1 \pm e^{- E_{3}^{+}(s) / T}} \right)}
    \int dt \sum_{\rm spin} |{\cal M}_{1 + 2 \to \axino + 3}|^{2} \,.
\end{align}

This is a particularly convenient form, because the integration variables, $s$ and $t$, are Lorentz-invariant.
We usually need to treat carefully the delta function that corresponds to the energy conservation, which results in a constraint on the phase space integration.
In Eq.~\eqref{eq:col-3}, such a constraint is automatically included in the energy $E_{3}^{\pm}$ that is a function of $s$ and $p_{\axino}$.
Therefore, what we need to do is to carefully calculate $E_{3}^{\pm}$ and squared matrix elements.

\subsection{Collision terms for specific processes}
\label{sec:deta-coll-term}

We summarize the collision terms and kinematic variables ($E^{\pm}$) used for specific axino production processes.
We also derive the Boltzmann equation for the number density by integrating over the axino phase space.
\begin{figure}
  \centering
  \includegraphics[width=0.6\linewidth]{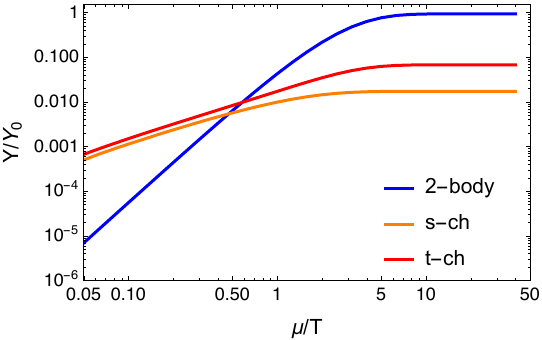}
  \caption{Time evolution of freeze-in axino yields. 
  Each line corresponds to the yield from each process, normalized by the total yield at present.}
  \label{fig:y-t}
\end{figure}
In Fig.~\ref{fig:y-t}, we show the evolution of the axino yields from Higgsino decay and scattering.
We calculate each yield by integrating Eqs.~\eqref{eq:n-2body}, \eqref{eq:n-tch}, and \eqref{eq:n-sch}, and it is normalized by the total axino yield at present $Y_{0}$.
Freeze-in ends at $T \sim 5 \mu$.
We can see that due to a large phase space factor, the total axino abundance is dominated by Higgsino 2-body decay.

\subsubsection{Higgsino 2-body decay}
\label{sec:2-body-decay-app}

When the Higgsino is the NLSP, Higgsino 2-body decay into the axino and lighter Higgs, ${\widetilde H} \to \axino + H_{L}$, is a dominant source of axinos.
In this case,
\begin{align}
  \label{eq:2body-epm-app}
  E_{2}^{+} = \infty \,, \qquad E_{2}^{-} = \frac{\mu^{2}}{4 p_{\axino}} \,,
\end{align}
and the collision term becomes
\begin{align}
  \label{eq:col-2body-2}
  \frac{g_{\axino}}{E_{\axino}} C_{{\widetilde H} \to \axino + H_{L}}(t, p_{\axino})
  &=
    \frac{T}{16 \pi p_{\axino}^{2}} \sum_{\rm spin} {\big |}{\cal M}_{{\widetilde H} \to \axino + H_{L}}{\Big |}^{2}
    \ln{\left[1+ \exp{\left( - \frac{\mu^{2}}{4 p_{\axino} T} - \frac{p_{\axino}}{T} \right)} \right]} \,.
\end{align}

The Boltzmann equation for the number density is obtained as
\begin{align}
  \label{eq:n-2body}
  \frac{dn_{\axino}}{dt} + 3 H n_{\axino}
  &=
    \int \frac{d^{3} p_{\axino}}{(2 \pi)^{3}} \frac{g_{\axino}}{E_{\axino}} C_{{\widetilde H} \to \axino + H_{L}}(t, p_{\axino}) \notag \\
  &=
    \frac{T}{32 \pi^{3}} \sum_{\rm spin} {\Big |} {\cal M}_{{\widetilde H} \to \axino + H_{L}}{\Big |}^{2}
    \int_{0}^{\infty} dp_{\axino} \ln{\left[ 1+ \exp{ \left(-\frac{\mu^{2}}{4 p_{\axino} T} - \frac{p_{\axino}}{T} \right)} \right]} \,.
\end{align}
When we use the Boltzmann distribution instead of the Fermi-Dirac distribution, $\ln \left[ 1+ \exp \left( - \dfrac{\mu^{2}}{4 p_{\axino} T} - \dfrac{p_{\axino}}{T} \right) \right] \simeq \exp \left( - \dfrac{\mu^{2}}{4 p_{\axino} T} - \dfrac{p_{\axino}}{T} \right)$, we obtain a simple formula,
\begin{align}
  \label{eq:n-2body-form}
  \frac{dn_{\axino}}{dt} + 3 H n_{\axino}
  &\simeq
    \frac{T \mu}{32 \pi^{3}} \sum_{\rm spin} {\Big |} {\cal M}_{{\widetilde H} \to \axino + H_{L}}{\Big |}^{2}
    K_{1} \left( \frac{\mu}{T} \right) \,.
\end{align}

\subsubsection{Higgs 2-body decay}
\label{sec:higgs-2-body}

We consider 2-body decay of the Higgses, $H_{L, H} \to \axino + {\widetilde H}$. 
Since the Higgsino mass is taken into account, we use the formula given in Sec.~\ref{sec:2-body-decay} as they are.

\subsubsection{Higgsino $s$-channel scattering}
\label{sec:s-channel-scattering}

For $s$-channel scattering, we consider $Q_{L}^{c} + t_{R} \to \axino + {\widetilde H}$. 
The extension to, {\it e.g.}, $H_{L} + H_{L}^{*} \to \axino + {\widetilde W}$ is straightforward.
In this case,
\begin{align}
  E_{3}^{+} &= \infty \,, \qquad E_{3}^{-} = \frac{s - \mu^{2}}{4 p_{\axino}} + \frac{\mu^{2} p_{\axino}}{s - \mu^{2}} \,.
\end{align}
The collision term becomes
\begin{align}
  \label{eq:s-col-2}
  \frac{g_{\axino}}{E_{\axino}} C_{Q_{L}^{c} + t_{R} \to \axino + {\widetilde H}}(t, p_{\axino})
  =&
    \frac{T}{256\pi^{3} p_{\axino}^{2}}
    e^{-p_{\axino} / T}
    \int_{\mu^{2}}^{\infty} ds \frac{1}{s-\mu^{2}}
    \ln\left( 1+ e^{- E_{3}^{-} / T} \right)\nonumber\\
    &\times
    \int_{- s + \mu^{2}}^{0} dt \sum_{\rm spin} {\Big |} {\cal M}_{Q_{L}^{c} + t_{R} \to \axino + {\widetilde H}}{\Big |}^{2} \,.
\end{align}

The Boltzmann equation for the number density is obtained as
\begin{align}
  \label{eq:n-sch}
  \frac{dn_{\axino}}{dt} + 3 H n_{\axino}
  =&
    \frac{T}{512 \pi^{5}}
    \int_{\mu^{2}}^{\infty} ds \frac{1}{s-\mu^{2}}
    \int_{- s + \mu^{2}}^{0} dt \sum_{\rm spin} {\Big |} {\cal M}_{Q_{L}^{c} + t_{R} \to \axino + {\widetilde H}}{\Big |}^{2}\nonumber\\
   & \times
    \int_{0}^{\infty} dp_{\axino} e^{-p_{\axino} / T} \ln\left( 1+ e^{- E_{3}^{-} / T} \right) \,.
\end{align}
Taking $\ln( 1+ e^{- E_{3}^{-} / T} ) \simeq  e^{- E_{3}^{-} / T}$, we obtain the same expression as in Eq.~\eqref{eq:n-tch-form} with only the squared matrix element being replaced.

\subsubsection{Higgsino $t$-channel scattering}
\label{sec:t-channel-scattering}

For $t$-channel scattering of the Higgsino, we consider ${\widetilde H} +Q_{L} \to \axino + t_{R}$.
The extension to other $t$-channel processes such as ${\widetilde H} + t_{R}^{c} \to {\tilde a} +Q_{L}^{c}$ and ${\widetilde W} + H_{L} \to \axino + H_{L}$ is straightforward.
In this case,
\begin{align}
  \label{eq:p-and-z-t}
  E_{3}^{+} = p_{3}^{+} = \infty \,, \qquad E_{3}^{-} = p_{3}^{-} = \frac{s}{4 p_{\axino}}.
\end{align}
One can obtain the collision term for other $t$-channel processes such as ${\widetilde H} + t_{R}^{c} \to \axino + Q_{L}^{c}$ by replacing the squared matrix element.
The collision term becomes
\begin{align}
  \label{eq:t-col-2}
  \frac{g_{\axino}}{E_{\axino}} C_{{\widetilde H} + Q_{L} \to \axino + t_{R}}(t, p_{\axino})
  =&
    \frac{T}{256\pi^{3} p_{\axino}^{2}}
    e^{-p_{\axino} / T}
    \int_{\mu^{2}}^{\infty} ds \frac{1}{s}
    \ln{\left[ 1 + \exp{\left(-\frac{s}{4 p_{\axino} T}\right)} \right]}\nonumber\\
   &\times \int_{-s+\mu^{2}}^{0} dt \sum_{\rm spin} {\Big |} {\cal M}_{{\widetilde H} + Q_{L} \to \axino + t_{R}}{\Big |}^{2} \,.
\end{align}

The Boltzmann equation for the number density is obtained as
\begin{align}
  \label{eq:n-tch}
  \frac{dn_{\axino}}{dt} + 3 H n_{\axino}
  =&
    \frac{T}{512 \pi^{5}}
    \int_{\mu^{2}}^{\infty} ds \frac{1}{s}
    \int_{-s+\mu^{2}}^{0} dt \sum_{\rm spin} {\Big |} {\cal M}_{{\widetilde H} + Q_{L} \to \axino + t_{R}}{\Big |}^{2}\nonumber\\
    &\times
    \int_{0}^{\infty} dp_{\axino} e^{-p_{\axino} / T}\ln{\left[ 1 + \exp{\left(-\frac{s}{4 p_{\axino} T}\right)} \right]} \,.
\end{align}
Taking $\ln \left[ 1+ \exp \left( {-\dfrac{s}{4 p_{\axino} T}} \right) \right] \simeq \exp \left(- \dfrac{s}{4 p_{\axino} T} \right)$, we obtain
\begin{align}
  \label{eq:n-tch-form}
  \frac{dn_{\axino}}{dt} + 3 H n_{\axino}
  &\simeq
    \frac{T}{512 \pi^{5}}
    \int_{\mu^{2}}^{\infty} ds \frac{1}{\sqrt{s}}
    K_{1}\left( \frac{\sqrt{s}}{T} \right)
    \int_{- s + \mu^{2}}^{0} dt \sum_{\rm spin} {\Big |} {\cal M}_{{\widetilde H} + Q_{L} \to \axino + t_{R}}{\Big |}^{2} \,,
\end{align}
which is concordant with the result of Ref.~\cite{Hall:2009bx}.

\subsubsection{Wino 3-body decay}
\label{sec:3-body-decay-app}

We consider ${\widetilde W} \to \axino + H_{L} + H_{L}^{*}$, where $H_{L}$ is massless.
In this case,
\begin{align}
  \label{eq:epm}
  E_{1}^{+} &= \infty \,, \qquad E_{1}^{-} = \frac{M_{2}^{2}-m_{2 3}^{2}}{4 p_{\axino}} + \frac{M_{2}^{2}p_{\axino}}{M_{2}^{2}-m_{2 3}^{2}} \,.
\end{align}
The collision term is
\begin{align}
  \label{eq:col-3body-2}
  \frac{g_{\axino}}{E_{\axino}} C_{{\widetilde W} \to \axino + H_{L} + H_{L}^{*}}(t, p_{\axino})
  =&
    \frac{T}{256\pi^{3} p_{\axino}^{2}}
    \int_{0}^{M_{2}^{2}} dm_{2 3}^{2} \frac{1}{M_{2}^{2}-m_{2 3}^{2}}
    \ln{\left(1+ e^{- E_{1}^{-} / T} \right)}\nonumber\\
    &\times
    \int_{0}^{M_{2}^{2} - m_{2 3}^{2}} dm_{2 \axino}^{2} \sum_{\rm spin} {\Big |} {\cal M}_{{\widetilde W} \to \axino + H_{L} + H_{L}^{*}}{\Big |}^{2} \,.
\end{align}

The Boltzmann equation for the number density is obtained as
\begin{align}
  \label{eq:n-3body}
  \frac{dn_{\axino}}{dt} + 3 H n_{\axino}
  =&
    \frac{T}{512 \pi^{5}}
    \int_{0}^{M_{2}^{2}} dm_{2 3}^{2} \frac{1}{M_{2}^{2}-m_{2 3}^{2}}
    \int_{0}^{M_{2}^{2} - m_{2 3}^{2}} dm_{2 \axino}^{2} \sum_{\rm spin} {\Big |} {\cal M}_{{\widetilde W} \to \axino + H_{L} + H_{L}^{*}}{\Big |}^{2}\nonumber\\
    &\times
    \int_{0}^{\infty} \ln\left(1+ e^{- E_{1}^{-} / T} \right) \,.
\end{align}
Taking $\ln \left(1+ e^{- E_{1}^{-} / T} \right) \simeq e^{- E_{1}^{-} / T}$, we obtain
\begin{align}
  \label{eq:n-3body-form}
  \frac{dn_{\axino}}{dt} + 3 H n_{\axino}
  &\simeq
    \frac{T}{512 \pi^{5}M_{2}}
     K_{1}\left(\frac{M_{2}}{T}\right)
    \int_{0}^{M_{2}^{2}} dm_{2 3}^{2}
    \int_{0}^{M_{2}^{2} - m_{2 3}^{2}} dm_{2 \axino}^{2} \sum_{\rm spin} {\Big |} {\cal M}_{{\widetilde W} \to \axino + H_{L} + H_{L}^{*}}{\Big |}^{2} \,.
\end{align}

\subsection{On the approximation of $1 \pm f^{\rm eq} \simeq 1$}
\label{sec:appr-1pm-fsim}

Here we analyze the validity of the approximation of $1 \pm f^{\rm eq} \simeq 1$.
In the above two sections, we obtain simple formulas of the collision term by ignoring these Bose-enhancement and Pauli-blocking terms.
Freeze-in becomes efficient at $m / T \sim O(1)$ ($m$ is the threshold scale), and it is not clear whether we can assume that $1 \pm f^{\rm eq} \simeq 1$ or not.

\begin{figure}
  \centering
  \begin{minipage}{0.5\linewidth}
    \includegraphics[width=1.0\linewidth]{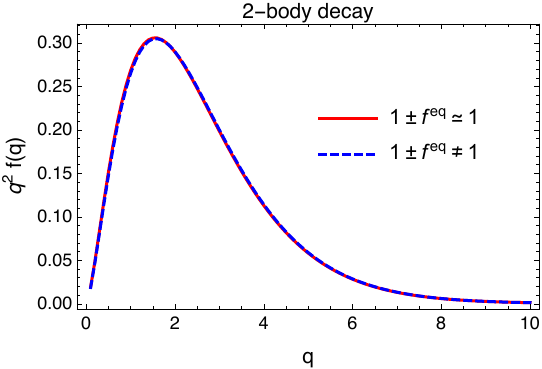}
  \end{minipage}
  \begin{minipage}{0.5\linewidth}
    \includegraphics[width=1.0\linewidth]{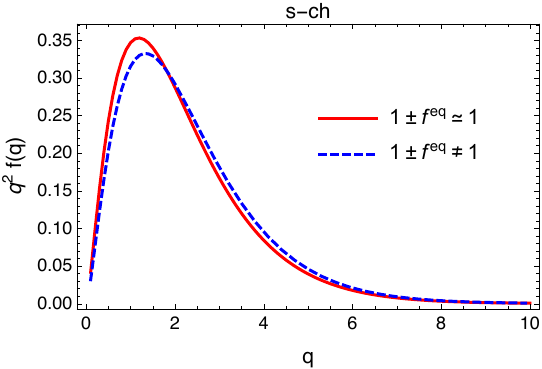}
  \end{minipage}%
  \begin{minipage}{0.5\linewidth}
    \includegraphics[width=1.0\linewidth]{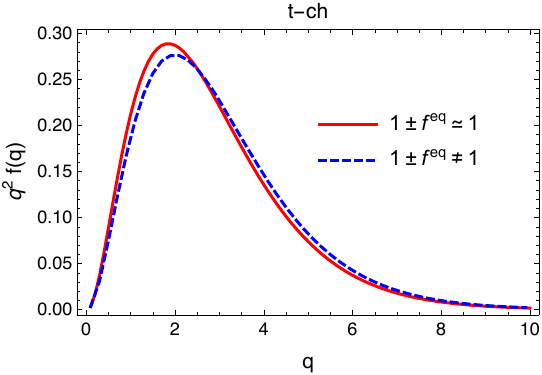}
  \end{minipage}
  \caption{Comparison of the phase space distributions with (solid) and without (dashed) the approximation of $1\pm f^{\rm eq} \simeq 1$ .
    The top panel shows the axino phase space distributions from Higgsino 2-body decay, the bottom left one shows those from $s$-channel scattering, and the bottom right one shows those from $t$-channel scattering.
  Each phase space distribution is normalized to give $\int dq q^{2}f(q) = 1$.}
  \label{fig:check-approx}
\end{figure}

With the term of $1 \pm f^{\rm eq}$, we have to rely on a brute-force calculation.
As an example, we assume that the Higgsino is the NLSP and axinos are produced by its decay and scattering.
Figure~\ref{fig:check-approx} compares the resultant phase space distributions with and without the approximation of $1 \pm f^{\rm eq} \simeq 1$ for Higgsino 2-body decay, $s$-channel scattering, and $t$-channel scattering.
For Higgsino 2-body decay, the approximation hardly affects the resultant phase space distribution.
On the other hand, there are slight discrepancies between the phase space distributions from scattering.
In terms of ${\tilde \sigma}$, the exact computations provide ${\tilde \sigma} = 2.72$ for $s$-channel scattering and ${\tilde \sigma} = 3.37$ for $t$-channel scattering, which show a few \% deviations from those in Table~\ref{tab:sigma-fit}.
Therefore, our computation in Sec.~\ref{sec:mom-dist-axino} entails the uncertainty of this level.

In the following, we list the collision terms used in the comparison of the axino phase space distributions.
Here we leave the axino mass non-zero just for generality of the formulas, but we take it to be zero in practice.
\begin{itemize}
\item For 2-body decay of $1 \to \axino + 2$, we consider the case where particle 1 is a fermion and particle 2 is a boson, as in Higgsino 2-body decay.
  The collision term is written as
\begin{align}
  \label{eq:col-2body-exact}
  \frac{g_{\axino}}{E_{\axino}} C_{1 \to \axino + 2}(t, p_{\axino})
  &=
    \frac{T}{16 \pi p_{\axino} E_{\axino}} \sum_{\rm spin} |{\cal M}_{1 \to \axino + 2}|^{2}
    \frac{g(E_{2}^{+}) - g(E_{2}^{-})}{e^{E_{\axino} / T}+1} \,,
\end{align}
where 
\begin{equation}
  \label{eq:gx}
  g(E) = \ln \left( 1 - e^{E / T} \right) - \ln \left( 1 + e^{(E + E_{\axino}) / T} \right) \,.
\end{equation}
The kinematic variables $E_{2}^{\pm}$ are obtained from Eq.~\eqref{eq:2body-epm}.
\item For scattering of $1 + 2 \to \axino + 3$, we consider the case where particles 1, 2, and 3 are fermions, as in Higgsino scattering.
  The collision term is written as 
  \begin{align}
    \label{eq:col-scat-exact}
    \frac{g_{\axino}}{E_{\axino}}C_{1 + 2 \to \axino + 3}(t, p_{\axino})
    =&
      \frac{1}{(2 \pi)^{4}16 E_{\axino}}
      \int_{p_{\rm min}}^{\infty} p_{3}^{2} dp_{3}
      \int_{\cos \theta_{\rm min}}^{\cos \theta_{\rm max}} d\cos \theta_{3}
      \frac{1 - f_{3}^{\rm eq}(E_{3})}{E_{3 }|\vec{p}_{3} + \vec{p}_{\axino}|} \notag \\
     & \times \int_{E_{\rm min}}^{E_{\rm max}} dE_{1} f_{1}^{\rm eq}(E_{1}) f_{2}^{\rm eq}(- E_{1} + E_{3} + E_{\axino})
      \int_{0}^{2 \pi} d\phi_{1} \sum_{\rm spin} |{\cal M}|^{2} \,.
  \end{align}
  The angle $\theta_{3}$ is defined as the polar angle of the three-momentum of particle 3 when that of the axino is regarded as the polar direction.
  The angles, $\phi_{1}$ and $\theta_{1}$, are defined, respectively, as the azimuthal and polar angles of the three-momentum of particle 1 
  when the sum of the three-momenta of the axino and particle 3 is regarded as the polar direction.
  The integration with respect to $\phi_{1}$ corresponds to that with respect to $t$.
  The energy-conservation delta function $\delta(E_{1} + E_{2} - E_{3} - E_{\axino})$ is used to fix $\cos \theta_{1}$ and also constrains the integration regions of $p_{3}$, $\cos \theta_{3}$, and $E_{1}$.
\end{itemize}

\section{Squared matrix elements}
\label{sec:matrix-elements}

In this appendix, we summarize the squared matrix elements used in the calculation.
For notational simplicity, in the following, we do not take the sum over particle degrees of freedom such as color and isospin other than spin.
One has to multiply them properly when plugging the squared matrix elements into the collision term.

\subsection{Higgsino}
\label{sec:higgsino}

The Higgsino forms a Dirac fermion ${\widetilde H}_\alpha = ({\widetilde H}^{+}, {\widetilde H}^{0})$, where $\alpha=1,2$ denotes the SU(2) indices.
We consider only the lighter Higgs $H_{L}$.
\begin{itemize}
\item  ${\widetilde H} \to \axino + H_{L}$
  \begin{equation}
  \label{eq:mat-2body}
  \sum_{\rm spin} |{\cal M}|^{2} = \mu^{2} \left(c_{H} \frac{\mu}{v_{PQ}} \right)^{2} \,.
\end{equation}

\item $t_{R} + Q_{L \beta}^{c} \to \axino + {\widetilde H}_{\alpha}$
  \begin{equation}
  \label{eq:mat-s-ch}
  \sum_{\rm spin} |{\cal M}|^{2} = |\epsilon_{\alpha \beta}| \left(c_{H} \frac{\mu}{v_{PQ}}\right)^{2}Y_{t}^{2}\frac{s-\mu^{2}}{s} \,.
\end{equation}

\item ${\widetilde H}_{\alpha} + t_{R}^{c} \to \axino + Q_{L \beta}^{c}$ and ${\widetilde H}_{\alpha} + Q_{L \beta} \to \axino + t_{R}$
  \begin{equation}
  \label{eq:t-ch}
  \sum_{\rm spin} |{\cal M}|^{2} = |\epsilon_{\alpha \beta}|\left(c_{H} \frac{\mu}{v_{PQ}} \right)^{2} Y_{t}^{2} \frac{t (t - \mu^{2})}{(t -m_{H_{L}}^{2})^{2}} \,,
\end{equation}
where the Higgs thermal mass, $m_{H_{L}} \simeq Y_{t} T / 2$, is introduced to avoid infrared divergence.
\end{itemize}

\subsection{Higgses}
\label{sec:higgses}

We consider the MSSM Higgses.
Diagonalizing the mass matrix of $H_{u}$ and $H_{d}$, we obtain the mass eigenstates, $H_{L}$ and $H_{H}$, as
\begin{equation}
  \label{eq:higgs-diag}
  \begin{pmatrix}
    H_{u\alpha} \\ -\epsilon_{\alpha\beta}H_d^{\beta*}
  \end{pmatrix}
  =
  \begin{pmatrix}
    \cos\alpha & \sin\alpha \\
    -\sin\alpha & \cos\alpha
  \end{pmatrix}
  \begin{pmatrix}
    H_{L \alpha} \\
    H_{H \alpha}
  \end{pmatrix}.
\end{equation}
\begin{itemize}
\item $H_{L, H} \to \axino + {\widetilde H}$
  \begin{equation}
    \label{eq:higgs-decay}
    \sum_{\rm spin} |{\cal M}|^{2}
    = \left( c_{H}\frac{\mu}{v_{PQ}} \right)^{2} m_{H_{L, H}}^{2}
    \left( 1 - \frac{\mu^{2}}{m_{H_{L, H}}^{2}} \right) \,.
  \end{equation}
  
\item
  In this case, we also need to consider Higgsino scattering via the exchange of the Higgses.
  The processes are the same as for Eqs.~\eqref{eq:mat-s-ch} and \eqref{eq:t-ch}, while the Higgs masses are taken into account,
  \begin{eqnarray}
    \sum_{\rm spin} |{\cal M}|^{2} = \left\{
      \begin{array}{l}
        |\epsilon_{\alpha\beta}| \left(Y_{H_{L, H} Q t} c_{H} \dfrac{\mu}{v_{PQ}} \right)^{2}
    \dfrac{s (s - \mu^{2})}{\left( s - m_{H_{L, H}}^{2} \right)^{2} + m_{H_{L, H}}^{2} \Gamma_{H_{L, H}}^{2}} \,, \qquad \text{for $s$-channel,}
      \label{eq:sch-Higgs} \\
        |\epsilon_{\alpha\beta}| \left(Y_{H_{L, H} Qt} c_{H} \dfrac{\mu}{v_{PQ}} \right)^{2}
    \dfrac{t (t - \mu^{2})}{\left( t - m_{H_{L, H}}^{2} \right)^{2}} \,, \qquad \text{for $t$-channel.}
    \label{eq:tch-Higgs}
      \end{array} \right.
  \end{eqnarray}
  The Higgs couplings to the top quarks are $Y_{H_{L} Q t} = Y_{t} $ and $Y_{H_{H} Q t} = Y_{t}  \tan \alpha$.
  The decay width of the Higgses is approximated as 
  \begin{equation}
  \Gamma_{H_{L, H}} \simeq \frac{Y_{H_{L, H} Q t}^{2}}{16 \pi} m_{H_{L, H}} \,.
  \end{equation}
  In order to avoid the double counting, we have to subtract the Higgs pole contributions from $s$-channel scattering,
  \begin{equation}
    \label{eq:sch-pole}
    \sum_{\rm spin} |{\cal M}|^{2}_{\rm pole}
    = |\epsilon_{\alpha\beta}| \left( Y_{H_{L, H} Q t} c_{H} \frac{\mu}{v_{PQ}} \right)^{2}
    \frac{\pi \, m_{H_{L, H}}^{2} \left( m_{H_{L, H}}^{2} - \mu^{2} \right)}{m_{H_{L, H}} \Gamma_{H_{L, H}}} \delta \left( s - m_{H_{L, H}}^{2} \right) \,.    
  \end{equation}
\end{itemize}

\subsection{Wino}
\label{sec:wino}

When considering the wino contribution, we assume that the Higgsino is heavier than the wino.
We consider wino 2-to-2 scattering and 3-body decay.
\begin{itemize}
\item ${\widetilde W}^{a} + H_{L \alpha} \to \axino + H_{L \beta}$
  \begin{equation}
    \label{eq:mat-wino-scat1}
    \sum_{\rm spin} |{\cal M}_{1} + {\cal M}_{2}|^{2} \,,
  \end{equation}
  where
  \begin{align}
  \sum_{\rm spin} |{\cal M}_{1}|^{2}
  =& 2 g_{2}^{2}(T^{a})_{\alpha}^{\beta} (T^{a})_{\alpha}^{\beta *}
    \left( c_{H} \frac{\mu}{v_{PQ}} \right)^{2} \frac{1}{(s - \mu^{2})^{2} + \mu^{2} \Gamma_{{\widetilde H}}^{2}}\notag \\
    & \times 
      \left[\mu^{2} (M_{2}^{2} - t) + 2 s_{\beta} c_{\beta} M_{2} \mu \left( s - m_{H_{L}}^{2} \right)\right.\nonumber\\
      &\left.+ 2 s_{\beta}^{2} c_{\beta}^{2} \left( - s u + s M_{2}^{2} + \mu^{2} t - M_{2}^{2} \mu^{2} - M_{2}^{2} m_{H_{L}}^{2} + m_{H_{L}}^{4} \right)\right]
      \label{eq:mat-wino1} \,,
  \end{align}
 \begin{align}
  \sum_{\rm spin} |{\cal M}_{2}|^{2}
  =& 2 g_{2}^{2} (T^{a})_{\alpha}^{\beta} (T^{a})_{\alpha}^{\beta *}
    \left( c_{H} \frac{\mu}{v_{PQ}} \right)^{2}\frac{1}{\left( u - \mu^{2} \right)^{2}} \notag \\
  & \times
   \left[\mu^{2}(M_{2}^{2}-t) + 2s_{\beta}c_{\beta}M_{2} \mu \left( u - m_{H_{L}}^{2} \right)\right.\nonumber\\
  &\left.  + 2 s_{\beta}^{2} c_{\beta}^{2} \left( - s u + u M_{2}^{2} + \mu^{2} t - M_{2}^{2} \mu^{2} - M_{2}^{2} m_{H_{L}}^{2} + m_{H_{L}}^{4} \right)\right]
    \,,
    \label{eq:mat-wino2}
  \end{align}
  and
  \begin{align}
  \sum_{\rm spin} {\cal M}_{1} {\cal M}_{2}^{*}
  =& - 2 g_{2}^{2} (T^{a})_{\alpha}^{\beta} (T^{a})_{\alpha}^{\beta *}
    \left( c_{H} \frac{\mu}{v_{PQ}} \right)^{2} \notag \\
  & \times
    \frac{s_{\beta} c_{\beta} M_{2} \mu \left( M_{2}^{2} - t \right) - 2 s_{\beta}^{2} c_{\beta}^{2} \left( s u - \mu^{2} t - M_{2}^{2} \mu^{2} - m_{H_{L}}^{4} \right)}{\left( (s - \mu^{2})^{2} + \mu^{2} \Gamma_{{\widetilde H}}^{2} \right) (u - \mu^{2})} \times (s - \mu^{2})
    \label{eq:mat-wino3} \,.
\end{align}
The decay width of the intermediate Higgsino is approximately given by
\begin{equation}
  \label{eq:higgsino-width}
  \Gamma_{{\widetilde H}} = \frac{3 g_{2}^{2} \mu}{16\pi}
  \left(1 + \frac{M_{2}^{2}}{\mu^{2}} - \frac{m_{H_{L}}^{2}}{\mu^{2}} + 4s_{\beta}c_{\beta}\frac{M_{2}}{\mu}\right)
  \sqrt{1 - \left(\frac{m_{H_{L}} + M_{2}}{\mu} \right)^{2}} \sqrt{1 - \left( \frac{m_{H_{L}} - M_{2}}{\mu} \right)^{2}} \,.
\end{equation}
The pole contribution has to be subtracted from $s$-channel scattering as in the case of Higgsino scattering (see eq.~\eqref{eq:sch-pole}).

\item $H_{L \alpha} + H_{L}^{* \beta} \to \axino + {\widetilde W}^{a}$

  The squared matrix element is obtained from Eq.~\eqref{eq:mat-wino-scat1} by replacing $t \leftrightarrow s$, taking $\Gamma_{{\widetilde H}} \to 0$, and multiplying $(-1)$ in the right hand sides of Eqs.~\eqref{eq:mat-wino1}--\eqref{eq:mat-wino3}.

\item ${\widetilde W}^{a} \to \axino + H_{L \alpha} + H_{L}^{* \beta}$

  The squared matrix element is obtained from Eq.~\eqref{eq:mat-wino-scat1} by the replacement of $s \to m_{3 \axino}^{2}$, $t \to m_{2 3}^{2}$, and $u \to m_{2 \axino}^{2}$ in the right hand sides of Eqs.~\eqref{eq:mat-wino1}--\eqref{eq:mat-wino3}.
\end{itemize}

\section{Fitting functions in the benchmark models}
\label{sec:fitt-fuct-benchm}

\begin{table}[h]
  \centering
  \begin{tabular}{lcccc} \hline
    & BM1 ($T_{R} > m_{H_{L}}$) & BM2 ($T_{R} = 1 \TeV$) & BM2 ($T_{R} = 100 \GeV$) & BM2 ($T_{R} = 50 \GeV$) \\ \hline
    a & 1.38 & 2.08 & 2.53 & 3.89 \\
    b & 1.26 & 0.991 & 0.903 & 0.904 \\ \hline
  \end{tabular}
  \caption{Fitting parameters for the realistic models. The fitting function takes the form of $q^{2} f(q) \propto q^{a} e^{- b q}$.}
  \label{tab:fit-bm}
\end{table}

We summarize the fitting functions of the phase space distribution in the realistic models in Sec.~\ref{sec:distr-realst-axino}.
In Table~\ref{tab:fit-bm}, the fitting parameters $a$ and $b$ are shown for the fitting function of $q^{2} f(q) \propto q^{a} e^{- b q}$.
Note that these parameters are highly model dependent since the phase space distribution is the superposition of the contributions from several processes.

\section{Note added on BM1}
\label{sec:on-bm1}

In Sec.~\ref{sec:higgsino-2-body}, we considered axino freeze-in production from Higgs 2-body decay, Higgsino $s$-channel scattering, and $t$-channel scattering.
In that discussion, we implicitly assumed that the lighter Higgs is {\it heavier} than Higgsino, more specifically, $m_{H_{L}} \simeq 1 \TeV$ and $\mu = 500 \GeV$.
However, sizable quantum corrections need to be taken into account for the lighter Higgs.
With the quantum corrections, the lighter Higgs doublet is {\it lighter} than Higgsino.
Thus, axinos are dominantly produced by 2-body decay of Higgsino into the axino and the lighter Higgs.
In the following, we repeat the analyses in Secs.~\ref{sec:higgsino-2-body} and \ref{sec:mps} with the corrected mass spectrum.

First, we calculate the axino phase space distribution.
In analyses of the early universe, care has to be taken for a thermal mass of the lighter Higgs doublet, $m_{H_{L}} \simeq y_{t} T / 2$ ($y_{t} \simeq 1$ is the top Yukawa coupling of the SM Higgs).
At a high temperature, $T \gtrsim 2 \mu / y_{t}$, the thermal mass makes the lighter doublet heavier than Higgsino, 
and thus the lighter Higgs decay into the axino and Higgsino occurs, instead of Higgsino decay into the axino and the lighter Higgs.
This effect does not substantially alter the resultant phase space distribution since the yield from Higgs decay during $T\gtrsim  2 \mu / y_{t}$ is subdominant compared to that from Higgsino decay during $T \lesssim 2 \mu / y_{t}$.
Another effect is that a sizable Higgs mass reduces the typical energy given to an axino.
This is similar to Higgs decay into the axino and Higgsino whose resultant phase space distributions are shown in Fig.~\ref{fig:dist-process}.
It argues that the larger daughter particle mass leads to the colder phase space distribution.
In the present case, a difference is that the daughter particle mass ({\it i.e.}, the lighter Higgs mass) is $T$-dependent.
The $T$-dependence is qualitatively different before and after electroweak symmetry breaking (EWSB), $T_{c} \simeq \sqrt{2} m_{h} / y_{t} \simeq 177 \GeV$.
The Higgs mass may be approximated by $m_{H_{L}} \simeq y_{t} \sqrt{T^2 - T_{c}^2} / 2$ for $T > T_{c}$, while may be negligible for $T < T_{c}$.

\begin{figure}
  \centering
  \includegraphics[width=0.7\textwidth]{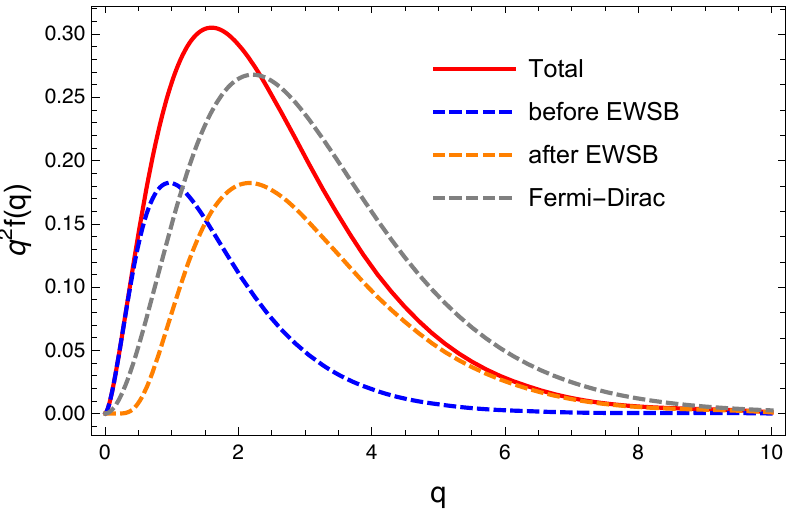}
  \caption{Axino phase space distribution in the BM1 scenario with the corrected mass spectrum. The total phase space distribution (red solid) consists of two contributions before (blue dashed) and after (orange dashed) EWSB and is to be compared with that (red solid) in the left panel of Fig.~\ref{fig:dist-real}. For comparison the Fermi-Dirac distribution (gray dashed) is also plotted.}
  \label{fig:dist-bm1-cor}
\end{figure}

Fig.~\ref{fig:dist-bm1-cor} shows the resultant axino phase space distribution from Higgsino decay into the axino and Higgs.
The phase space distribution is similar to that from Higgs decay into the axino and massless Higgsino (blue solid in the right panel of Fig.~\ref{fig:dist-process}).
It implies that the thermal mass does not make axino DM significantly colder.
Actually, the phase space distribution is hotter than that of the original mass spectrum where
the lighter Higgs is heavier than Higgsino (red line in the left panel of Fig.~\ref{fig:dist-real}).
In that case, a sizable Higgsino mass makes the axino DM colder.
In the present case, it turns out that the Higgs thermal mass does not play a sizable role.
The situation may change if we vary $y_{t}$ and $\mu$, while a systematic study is beyond the scope of this paper.
We refer to forbidden freeze-in DM~\cite{Dvorkin:2019zdi, Darme:2019wpd} for a different but illustrative setup, 
where a freeze-in process does not occur in vacuum but occurs in the thermal environment.
There the thermal mass significantly makes freeze-in DM colder~\cite{Dvorkin:2019zdi}.

Next, we calculate the linear matter power spectrum of $7 \keV$ axino DM from the obtained phase space distribution. 
The phase space distribution is well approximated by that from 2-body decay into massless particles, $q^{2} f(q) \propto q^{1.5} \exp(- q)$, as mentioned above.
\begin{figure}
  \centering
  \includegraphics[width=0.7\textwidth]{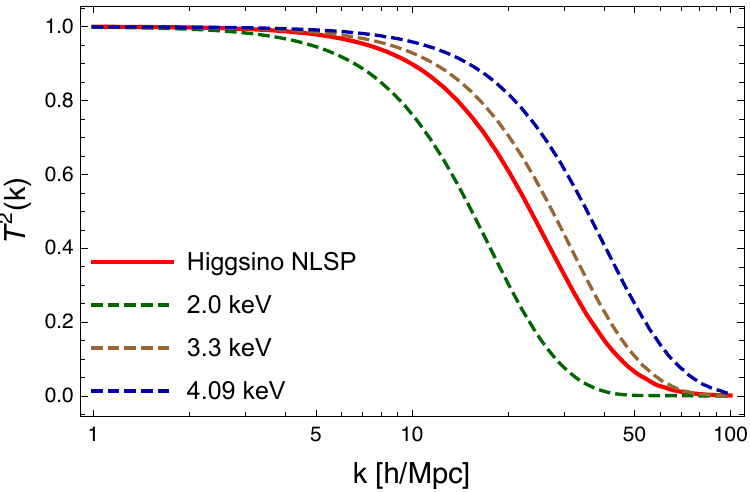}
  \caption{Squared transfer function in the BM1 scenario with the corrected mass spectrum. This squared transfer function (red solid) is to be compared with that (red solid) in the right panel of Fig.~\ref{fig:T2k}. The conventional WDM models with $m_{\rm WDM} = 2.0$, $3.3$, and $4.09 \keV$ are shown for comparison (dashed).}
  \label{fig:T2k-bm1-cor}
\end{figure}
Fig.~\ref{fig:T2k} shows the resultant squared transfer function ${\cal T}^{2} (k)$.
The squared transfer function in the present case (red solid line in Fig.~\ref{fig:T2k-bm1-cor}) is similar to that for 2-body decay into massless particles (red solid line in the left panel of Fig.~\ref{fig:T2k}) as expected.
The cutoff scale in BM1 with the corrected mass spectrum is smaller than that with the original mass spectrum (red solid line in the right panel of Fig.~\ref{fig:T2k}).
The corresponding WDM mass from Eq.~\eqref{eq:mass-rel} is $2.5 \keV ({\tilde \sigma} / 3.6)^{-3/4} = 2.9 \keV$ with the corrected mass spectrum, while is $3.6 \keV$ with the original mass spectrum.
The BM1 scenario with the original mass spectrum is consistent with the Ly-$\alpha$ forest constraint of $m_{\rm WDM} \gtrsim 3.3 \keV$.
In contrast, the BM1 scenario with the corrected mass spectrum is inconsistent with that of $m_{\rm WDM} \gtrsim 3.3 \keV$, while is still consistent with the Ly-$\alpha$ forest constraint of $m_{\rm WDM} \gtrsim 2.0 \keV$. 

Finally, we estimate the entropy dilution factor $\Delta$ required for $7 \keV$ axino DM to satisfy the Ly-$\alpha$ forest constraint of $m_{\rm WDM} \gtrsim 5.3 \keV$.
From Eq.~\eqref{eq:mass-rel-ent-pro}, $\Delta$ needs to be larger than 11.6 in BM1 with the corrected mass spectrum, while 4.7 with the original mass spectrum.
\begin{figure}
  \centering
  \includegraphics[width=0.7\textwidth]{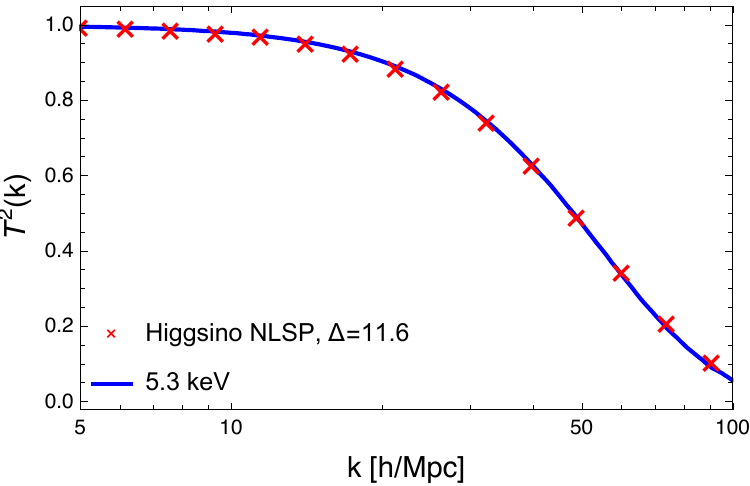}
  \caption{Squared transfer function in the BM1 scenario with the corrected mass spectrum. We introduce entropy production $\Delta$. This squared transfer function (red crosses) is to be compared with that (red triangles) in Fig.~\ref{fig:T2k-entropy}. The conventional WDM model with $m_{\rm WDM} = 5.3 \keV$ is shown for comparison (blue solid line).}
  \label{fig:T2k-entropy-cor}
\end{figure}
Fig.~\ref{fig:T2k-entropy-cor} confirms that the resultant matter power spectrum with $\Delta = 11.6$ is comparable with that of the conventional WDM model with $m_{\rm WDM} = 5.3 \keV$.



\bibliographystyle{utphys}
\bibliography{axino_wdm}

\end{document}